\documentclass[preprint,10pt]{elsarticle}

\usepackage{amssymb}
\usepackage{amsmath}
\newtheorem{example}{Example}
\usepackage{booktabs}   
\usepackage{xcolor,colortbl}
\usepackage{algorithm}
\usepackage{algpseudocode}
\usepackage{rotating}
\usepackage{tabularx} 
\usepackage{caption}
\usepackage{multirow}
\usepackage{listings}

\begin{document}

\begin{frontmatter}



\title{BridgeNet: A Hybrid, Physics-Informed Machine Learning Framework for Solving High-Dimensional Fokker-Planck Equations}

\author[1]{Elmira Mirzabeigi}
\ead{e.mirzabeigi@modares.ac.ir}

\author[1]{Rezvan Salehi}
\ead{r.salehi@modares.ac.ir}

\author[2,3]{Kourosh Parand\corref{cor1}}
\ead{k_parand@sbu.ac.ir}
\cortext[cor1]{Corresponding author}

\address[1]{Department of Applied Mathematics, Faculty of Mathematical Sciences, Tarbiat Modares University, Tehran, Iran}

\address[2]{Department of Computer and Data Sciences, Faculty of Mathematical Sciences, Shahid Beheshti University, Tehran, Iran}

\address[3]{Institute for Cognitive and Brain Sciences, Shahid Beheshti University,Tehran, Iran}

\begin{abstract}
BridgeNet is a novel hybrid framework that integrates convolutional neural networks with physics-informed neural networks to efficiently solve non-linear, high-dimensional Fokker–Planck equations (FPEs). Traditional PINNs, which typically rely on fully connected architectures, often struggle to capture complex spatial hierarchies and enforce intricate boundary conditions. In contrast, BridgeNet leverages adaptive CNN layers for effective local feature extraction and incorporates a dynamically weighted loss function that rigorously enforces physical constraints. Extensive numerical experiments across various test cases demonstrate that BridgeNet not only achieves significantly lower error metrics and faster convergence compared to conventional PINN approaches but also maintains robust stability in high-dimensional settings. This work represents a substantial advancement in computational physics, offering a scalable and accurate solution methodology with promising applications in fields ranging from financial mathematics to complex system dynamics.
\end{abstract}




\end{frontmatter}


\section{Introduction}
The advancement of mathematical modeling and computational methods has greatly accelerated progress in various scientific fields. These methods offer new perspectives and enhance our capability to forecast outcomes in physics, biology, economics, and engineering. Scientists use a variety of numerical techniques, from finite difference and meshless methods to the development of sophisticated knowledge-based potential functions, to solve complex differential equations and analyze data \cite{moayeri2023solving, rasanan2023numerical, mirzabeigi2023designing, eslahchi2021application, raj2024efficient, rasanan2024response,  marchetti2024fast}.
Among these, the Fokker-Planck equations (FPEs) stand out as a cornerstone in the study of stochastic processes, describing the time evolution of probability densities under the influence of diffusion and drift. These equations are pivotal in theoretical physics and emerging areas like financial math and complex system dynamics in cognitive science.

High-dimensional Fokker-Planck equations are a class of partial differential equations that describe the time evolution of probability density functions for stochastic processes, particularly in high-dimensional state spaces. These equations have a long history, originating in statistical mechanics, where they were first used to model the behavior of particles under random forces \cite{risken1996fokker,elgin1984fokker}. Over time, their application expanded to various fields, including biological systems, where they model population dynamics, and financial mathematics, where they help in option pricing and risk assessment. 

However, solving high-dimensional Fokker-Planck equations poses significant computational challenges due to the curse of dimensionality, which leads to exponential growth in the complexity of numerical methods. Traditional solvers, such as finite difference and finite element methods, struggle to efficiently handle the high dimensionality and intricate boundary conditions often present in such systems, necessitating the development of novel approaches that can provide both accuracy and computational feasibility.

Recent advancements in machine learning, particularly deep learning, have created novel avenues for addressing these challenges \cite{karami2023comparison}. Neural networks, when applied to solving equations, leverage their ability to learn and model complex relationships, providing a flexible and powerful approach to approximate solutions with high accuracy across diverse mathematical problems \cite{wojtkiewicz2000numerical, panju2020symbolically, omidi2022learning, hajimohammadi2021fractional, sirignano2018dgm, lu2022multifidelity, babaei2022jdnn, hajimohammadi2021legendre, moayeri2020dynamical, hadian2020single, rad2017meshfree, rabiei2020collocation, parand2020solving, hajiollow2021recovering, lotfi2021numerical, moayeri2022efficient, baharifard2022novel, parand2022least, asghari2023fpga, hajimohammadi2023novel}. Physics-informed neural networks (PINNs) have emerged as a promising approach, integrating physical laws directly into the learning mechanisms of deep neural networks. PINNs leverage the structure of neural networks to approximate the solutions of differential equations, ensuring that these solutions are consistent with physical laws by incorporating the differential equation as part of the loss function during training. This integration not only enforces adherence to physical principles but also enhances the efficiency and accuracy of the neural network in scenarios where traditional numerical methods might struggle, such as problems with complex geometries or boundary conditions \cite{raissi2019physics, meng2020composite, shin2020convergence, hu2021extended, namaki2023use, jagtap2022physics, liu2024discontinuity, donnelly2024physics, lu2024physics}. Various enhancements to the PINN approach, such as integrating adaptive activation functions, employing multi-fidelity modeling, and incorporating uncertainty quantification, significantly refine its effectiveness and broaden its applicability across different scientific and engineering domains \cite{babaei2024solving, mazraeh2024gepinn, zhang2024stochastic, lahariya2021physics, haitsiukevich2023improved, rafiq2022dsfa, zobeiry2021physics, pan2024ro, lee2024anti, eshkofti2024new, xiang2022self, psaros2022meta, habib2022developing, yang2022multi}. Despite their advantages, PINNs often require extensive computational resources and may struggle with spatial data representations typical in image and signal processing domains. On the other hand, CNNs are also famous for their efficiency in handling spatial hierarchy in data through their inherent translation invariance and local connectivity properties \cite{krizhevsky2012imagenet}. By leveraging CNNs' ability to extract and process spatial features, we hypothesize that a synergistic combination with PINNs could yield a robust framework for solving PDEs more efficiently.

Unlike conventional numerical solvers and standard PINNs that rely solely on heuristic adjustments or brute-force computations, BridgeNet presents an innovative hybrid framework that combines convolutional neural networks' spatial feature extraction abilities with the rigor of physics-informed constraints. Our contributions can be summarized as follows:

\begin{itemize}
	\item \textbf{\small Innovative Hybrid Architecture:} BridgeNet is the first framework to integrate CNN-based spatial processing with physics-informed constraints, effectively capturing localized dependencies while strictly enforcing the governing physical laws.
	\item \textbf{\small Enhanced Accuracy and Stability:} Extensive evaluations across one-dimensional, multi-dimensional, and non-linear Fokker–Planck equations demonstrate that BridgeNet achieves significantly lower error metrics (e.g., MSE, MAE, \(L_\infty\)) and maintains stable convergence even under complex boundary conditions.
	\item \textbf{\small Efficient Computational Performance:} By leveraging adaptive hyperparameter tuning and efficient CNN representations, BridgeNet converges rapidly with fewer epochs, mitigating the computational burden typically associated with high-dimensional problems.
	\item \textbf{\small Broad Impact in Computational Physics:} This novel fusion of deep learning with physics-based modeling not only sets a new benchmark for solving differential equations but also paves the way for addressing a wide range of real-world, high-dimensional, and stiff PDE challenges.
\end{itemize}

This study is structured as follows: Section 2 provides an overview of the theoretical background of FPEs and the exponential growth equation; Section 3 details the architecture and training of BridgeNet; Section 4 presents the experimental results; and the final section discusses the research implications and outlines potential future directions.
\section{Background and Related Work}
The exploration of differential equations, particularly Fokker-Planck equations (FPEs) and exponential growth model, has been intensive due to their broad applicability across physics, finance, biology, and more. This section discusses seminal works and recent advancements in these equations' numerical and analytical solutions, focusing on integrating traditional methods with contemporary machine-learning techniques.
\subsection{Fokker-Planck Equations: From Classical to Contemporary Approaches}
In the early 1990s, a group of researchers led by Palleschi delved into studying Fokker-Planck equations \cite{palleschi1990numerical, palleschi1992numerical}. They proposed a swift and precise algorithm for the numerical resolution of equations resembling the Fokker-Planck equations. Around the same time, Vanaja introduced an iterative method for solving FPEs \cite{vanaja1992numerical}. Between 1990 and 2010, various numerical methods were employed to enhance the outcomes of these equations, each method bringing its unique nuances and considerations \cite{zorzano1999numerical, dehghan2006use, tatari2007application, lakestani2009numerical}. Also, in 2011, Kazem et al. explored two numerical meshless methods for solving the Fokker-Planck equations, utilizing radial basis functions and the collocation method, with one approach based on Kansa's method and the other on Hermite interpolation \cite{kazem2012radial}. In 2018, Parand et al. presented a new numerical approach for solving time-dependent linear and non-linear FPEs, utilizing the Crank-Nicolson method for time discretization and the Generalized Lagrange Jacobi Gauss-Lobatto collocation method for space discretization \cite{parand2018generalized}. 

In 2022, Czarnetzki et al. demonstrated an alternative approach to describing the interaction between electrons and electric fields in low-pressure plasmas, utilizing the Fokker-Planck equations in tandem with the Langevin equation, and emphasizing its application in the problem of combined Ohmic and stochastic heating in inductively coupled plasmas \cite{czarnetzki2022describing}. At the same time, Tabandeh et al. introduced a novel numerical method for solving the FPEs, which governs the uncertainty propagation in dynamical systems driven by stochastic processes, by employing physics-based mixture models and Bayesian inference to estimate unknown parameters, thereby facilitating the integration of system response data with the governing equation \cite{tabandeh2022numerical}. Also, Liu et al. presented the development and analysis of numerical methods for high-dimensional FPEs, utilizing generative models from deep learning and formulating the Fokker-Planck equations as a system of ordinary differential equations (ODEs) on a finite-dimensional parameter space \cite{liu2022neural}. In 2023, Boffi et al. explored an alternative method for integrating the time-dependent Fokker-Planck equations, which involves modeling the ‘score’ (the gradient of the logarithm of the solution) with a deep neural network that is trained on the fly, offering direct access to challenging quantities such as the probability current, the density, and its entropy \cite{boffi2023probability, chen2023random}. In 2024, Aszhar et al. used the Fokker-Planck equations to study perfectly damped Brownian motion within a harmonic potential, aiming to identify additional potentials that could expedite the system’s return to equilibrium \cite{aszhar2024method}. A physics-informed neural network (PINN) with layer-wise locally adaptive activation functions (L-LAAF) and a learning rate decay strategy is used by Zhang et al. to investigate the stochastic propagation of time-dependent FPEs \cite{zhang2024stochastic}. At the same time, in separate research, Wang et al., Wang et al., and Neena et al. have been solving these issues with different methods for different applications \cite{wang2024dynamical, wang2024tensor, neena2024nonstandard}.

High-dimensional Fokker-Planck equations have also been extensively studied using various approaches to handle the complexities associated with their solution. Zhang and Zhou (2013) proposed a sparse grid approach to address the curse of dimensionality \cite{zhang2013sparse}, while Sirignano and Spiliopoulos (2018) introduced the Deep Galerkin Method (DGM) to leverage deep learning for solving high-dimensional PDEs, including Fokker-Planck equations \cite{sirignano2018dgm}. Han et al. (2018) presented a deep learning framework specifically tailored to solve high-dimensional PDEs, demonstrating significant improvements in handling the computational challenges \cite{han2018solving}. Additionally, Raissi et al. (2019) introduced physics-informed neural networks (PINNs), which have been particularly effective in incorporating domain-specific knowledge into the solution process, significantly improving the accuracy of high-dimensional FPE solutions \cite{raissi2019physics}.
\subsection{Exponential Growth Equation: From Classical to Contemporary Approaches}
Around 1789, Thomas Robert Malthus published an article on the principle of population and introduced the Malthusian growth model, which became a simple exponential growth model. In this model, which represents exponential growth, the rate of increase is proportional to the current value. Moreover, Dewar (1993) proposed a model that intricately links the growth of roots and shoots through the interactions of carbon, nitrogen, and water transport mechanisms via Munch phloem flow. This model provides a framework for understanding plant physiology and exemplifies the exponential growth equation \cite{dewar1993root}.

Dehghan et al. (2012), in a paper introducing a numerical method for fractional equations, described techniques that can complement PINN approaches in managing time-dependent equations such as exponential growth under variable conditions \cite{saadatmandi2012sinc}.
Also, Yang et al. (2017) numerically solved equations such as exponential growth using two models with the expansion method of the Galerkin series \cite{li2017experimental}. Spiliopoulos et al. (2018) explored a deep learning-based algorithm that parallels PINNs in its approach to solving PDEs efficiently \cite{sirignano2018dgm}.
In 2019, Karniadakis et al. introduced PINNs and their applications across different differential equations, including those with exponential growth characteristics, showcasing their capability to handle complex nonlinear behaviors efficiently \cite{raissi2019physics}. In 2020, Karniadakis et al. improved the solution of high-dimensional differential equations involving exponential growth by comprehensively reviewing physics-informed neural networks (PINN), generalizing the convergence of this method, and using Extended PINNs (XPINN)\cite{meng2020composite, shin2020convergence, hu2021extended}.
Romano et al. (2022) explore multi-fidelity learning approaches within PINNs for rapid design iterations in complex systems, demonstrating the flexibility of PINNs in adapting to various types of equations, including growth dynamics in nanoscale systems \cite{lu2022multifidelity}.
\section{Methodology}
\begin{sidewaysfigure}
	\includegraphics[width=\columnwidth]{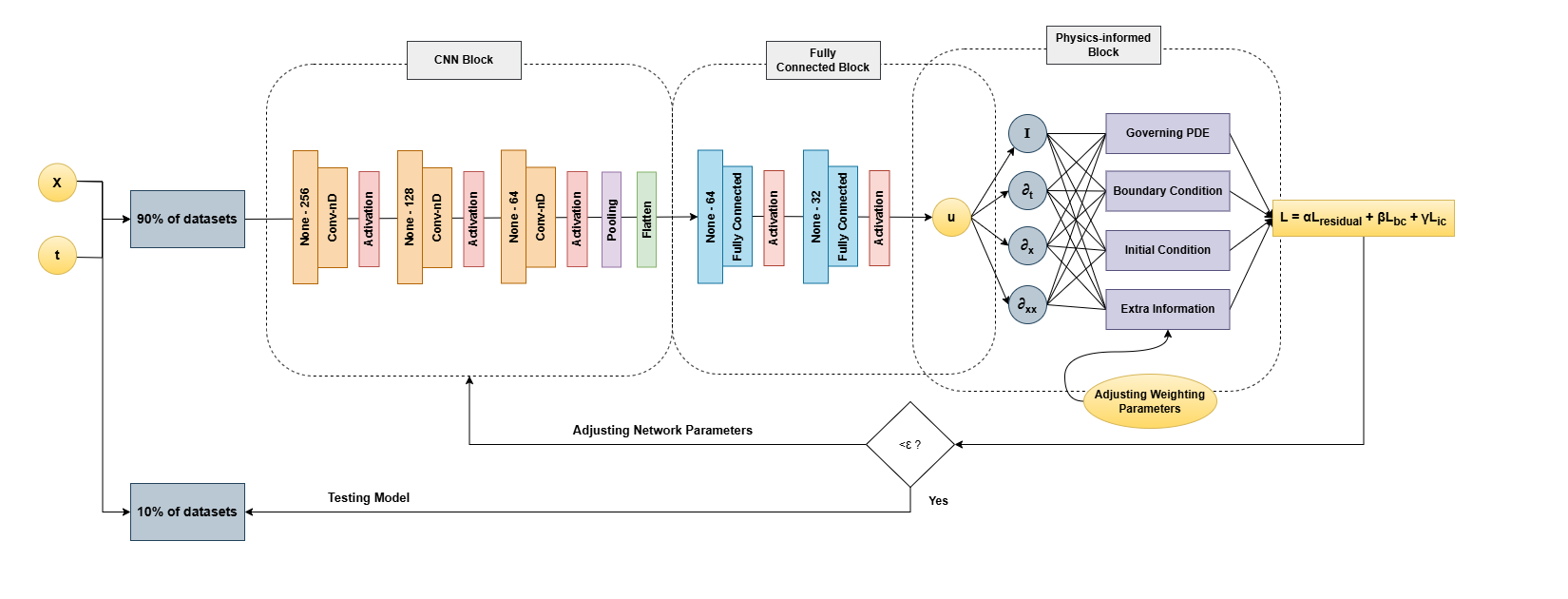}%
	\caption{A schematic representation of a \textbf{Convolutional Neural Network (CNN)} integrated with \textbf{physics-informed neural networks (PINNs)} for solving high-dimensional Fokker-Planck equations is illustrated. The input consists of a vector of spatial ($X$) and temporal ($t$) data, where $X$ is represented as an \textbf{n-1 dimensional matrix}. The data is split into training (90\%) and testing (10\%) datasets. The \textbf{CNN block} processes the input through a series of \textbf{convolutional and activation layers}, followed by a \textbf{fully connected block}. This block performs local input checks, extracting spatial features and capturing dependencies to enhance data representation. Notably, the use of \textbf{conv-nD} layers enables the CNN to process data in n-dimensional space, efficiently handling the \textbf{n-1 dimensional matrix} representation of the spatial data. The \textbf{physics-informed block} incorporates governing partial differential equations (PDEs), boundary conditions, initial conditions, and additional information, adjusting weighting parameters to minimize the total loss function: $L = \alpha L_{\text{residual}} + \beta L_{\text{bc}} + \gamma L_{\text{ic}}$, where $\alpha$, $\beta$, and $\gamma$ are weighting parameters. The iterative training continues until the model's error reaches a specified threshold.}
	\label{fig:diagram}
\end{sidewaysfigure}

The core objective of our research is to develop and implement BridgeNet, capable of solving linear and non-linear Fokker-Planck equations (FPEs). Integrating the physics-informed constraints within the convolutional neural network architecture enables our model to leverage spatial and temporal data effectively. The methodology section outlines the implementation steps, including network architecture, training procedure, and the integration of physical laws into the learning process. Figure \ref{fig:diagram} provides a schematic depiction of the BridgeNet developed for this research.
\subsection{Network Architecture}
Traditional PINNs primarily use fully connected Multi-Layer Perceptrons (MLPs) to approximate solutions to PDEs. However, MLPs treat spatial and temporal features independently, making them inefficient at capturing local spatial dependencies in high-dimensional PDEs such as Fokker-Planck equations. This results in slow convergence and limited generalization, especially in complex boundary conditions.

In contrast, Convolutional Neural Networks (CNNs) offer a superior approach by leveraging weight sharing, local receptive fields, and spatial hierarchies. These properties allow CNNs to extract localized patterns in PDE solutions while maintaining computational efficiency. By integrating CNN layers into the traditional PINN framework, we introduce BridgeNet, a hybrid model that enhances spatial feature extraction while preserving physics-informed constraints. The CNN component enables efficient learning of localized structures in PDE solutions, improving both accuracy and stability over standard PINNs. The following sub-sections describe architecture of BridgeNet, detailing its convolutional layers, activation functions, and integration of physics-informed constraints:
\subsubsection{Convolutional Layers}
The architecture of our network is dynamically adjusted based on the complexity and type of differential equations being solved. When addressing PDEs, particularly when the input comprises spatial $X$ and temporal $t$ components, where $X$ is represented as an (n-1)-dimensional matrix, the model leverages conv-nD layers to efficiently extract features from this high-dimensional input. These layers facilitate structured learning in n-dimensional space, preserving spatial coherence and capturing dependencies across different dimensions, thereby enhancing the model’s capability to process multi-dimensional data effectively. Across all configurations, the network architecture typically includes three convolutional stages that progressively refine the channel depth, from 256 channels in the initial layer and reducing to 128 and then 64 in the subsequent layers. This setup enables the network to extract and learn a diverse set of features from the temporal and spatial data, with all convolutional layers uniformly utilizing a kernel size of 3 and padding of 1, thus preserving the dimensionality of the input through each layer's transformations. In most scenarios, the described architecture is employed, though adjustments in layer depth and number of channels were made as dictated by the specific requirements of each problem set.
\subsubsection{Activation Functions}
Our network's activation functions are meticulously selected based on the specific requirements of the differential equations under study. For linear differential equations, we utilize the Rectified Linear Unit (ReLU) activation function, prized for its straightforwardness and effectiveness, particularly in scenarios where gradient vanishing is minimal. In contrast, for non-linear differential equations, our architecture incorporates a diverse array of sophisticated activation functions such as Tanh, Exponential Linear Unit (ELU), and SoftPlus. These are carefully chosen for their robust non-linear processing capabilities, which help mitigate issues related to vanishing and exploding gradients, thus substantially boosting the network's ability to discern and learn complex patterns within the data.

\subsubsection{Pooling Layer}
Following the convolutional layers, an adaptive average pooling layer (AdaptiveAvgPool1d) reduces the temporal dimension to a single value per feature map. This reduction is critical for summarizing the learned temporal features into a format suitable for the fully connected layers.
\subsubsection{Fully Connected Layers}
The pooled output is flattened and then processed through two fully connected layers. The first dense layer reduces the feature dimension from 64 to 32, followed by a reduction to a single output value representing the solution to the differential equation at a given point in time.
\subsubsection{Output}
The network's final output is a single scalar value per input sample, which directly corresponds to the solution of the differential equation at the specified time.
\subsection{Incorporating Physics-Informed Constraints}
Integrating differential equation constraints within neural networks, particularly in training BridgeNet, is a crucial focus of our research. Rather than embedding constraints directly into the network's architecture, we employ a dynamic loss function that embeds these constraints, ensuring model compliance with the governing physical laws. This loss function is crucial when dealing with partial differential equations (PDEs), as it adjusts based on the specific equation type and its requirements. In this study, our loss function effectively enforces the physical laws dictated by the Fokker-Planck equations through penalty terms for any deviations from these laws. This function comprises several components, each designed to ensure the model's predictions are accurate and faithful to the underlying dynamics of the addressed equations that are formulated as follows:
\begin{equation}
	L = \alpha L_{\text{residual}} + \beta L_{\text{bc}} \cdot \textbf{1}_{\text{PDE}} + \gamma L_{\text{ic}},
\end{equation}
where $\alpha$, $\beta$, and $\gamma$ are weighting parameters that manage the importance of each component, and $\textbf{1}_{\text{PDE}}$ is an indicator function that activates the boundary condition component only for PDE scenarios. In the following, we discussed these details:
\subsubsection{Residual Loss}
In our model, the residual function directly evaluates the consistency of the model outputs with the governing differential equations. Unlike traditional methods that measure discrepancies between predicted and expected behaviors, our approach directly substitutes the model's predictions into the differential equations to compute the residual. This method effectively assesses how well the physics-informed predictions satisfy the equations at specified data points. For PDEs, the loss component reflects how accurately the model adheres to the dynamics described by these equations. Depending on the equation's nature, involving spatial dimensions, the residual is mathematically formulated as:
\begin{equation}
	L_{\text{residual}} = \frac{1}{N_x, N_t} \sum_{i=1}^{N_x}\sum_{j=1}^{N_t} \left( f(u(x_i, t_j), x_i, t_j) \right)^2.
\end{equation}

Here, $f$ represents the differential operator tailored to each type of equation, applied to the neural network’s output $u$ at points specified by $x_i$ and $t_j$ for PDEs.
\subsubsection{Boundary and Initial Condition Loss}
This loss is particularly relevant for PDEs, where boundary conditions at the spatial domain's edges and initial conditions at the start of the temporal domain are essential:
\begin{equation}
	L_{\text{bc}} = \frac{1}{N_t} \sum_{j=1}^{N_t} \left( u(\text{bc}_j, t_j) - \text{BC}(\text{bc}_j, t_j) \right)^2,
\end{equation}
\begin{equation}
	L_{\text{ic}} = \frac{1}{N_x} \sum_{i=1}^{N_x} \left( u(x_i, t_0) - IC(x_i) \right)^2.
\end{equation}

$\text{BC}$ represents the boundary conditions, applied at specific spatial boundary points $\text{bc}_j$ and times $t_j$. Also, $IC$ denotes the initial condition applicable.
\subsubsection{Adjustment of the Weighting Parameters}
We use a smart greedy search algorithm to optimize the weighting parameters $\alpha$, $\beta$, and $\gamma$ in a loss function for solving differential equations using BridgeNet to adaptively update these parameters based on their sensitivity to the overall error reduction, thus achieving the most accurate solution. Algorithm \ref{Algorithm:1} begins with equal values for $\alpha$, $\beta$, and $\gamma$ and iteratively updates them by measuring the sensitivity of the total loss function to small changes in each parameter. At each iteration, only the parameter with the greatest impact on the loss (highest sensitivity) is updated, focusing the algorithm on the most critical part of the loss function. After each iteration, the parameters are clipped to be non-negative, and the total loss is recalculated, continuing this process until convergence is achieved. This results in the optimal values for $\alpha$, $\beta$, and $\gamma$, providing the most accurate solution.

The choice of a learning rate of 0.1 and 100 iterations balances convergence speed and solution accuracy. A learning rate of 0.1 ensures sufficiently large parameter updates without overshooting the minimum of the loss function, while smaller learning rates would slow convergence and require more iterations to reach an optimal solution. This rate allows meaningful progress in minimizing the loss while maintaining stability in updates. The use of 100 iterations is based on practical experimentation with similar problems, where this number is generally sufficient to converge to a near-optimal set of parameters, providing a reasonable balance between computational efficiency and solution quality. A learning rate and iterations offer a good compromise, ensuring fast convergence without compromising the stability of the search process, making these values well-suited for most BridgeNet problems.
\begin{algorithm}
	\caption{Smart Greedy Search for $\alpha$, $\beta$, $\gamma$}
	\label{Algorithm:1}
	\begin{algorithmic}[1]
		\State \textbf{Initialize:} $\alpha = 1.0$, $\beta = 1.0$, $\gamma = 1.0$, learning\_rate = 0.1, min\_loss = $\infty$, num\_iterations = 100
		\Function{compute\_total\_loss}{$\alpha$, $\beta$, $\gamma$}
		\State \textbf{return} $\alpha$ * compute\_residual\_loss(predictions) + $\beta$ * compute\_bc\_loss(predictions) $\cdot$ indicator\_PDE() + $\gamma$ * compute\_ic\_loss(predictions)
		\EndFunction
		\Function{compute\_sensitivity}{$\alpha$, $\beta$, $\gamma$}
		\State base\_loss = compute\_total\_loss($\alpha$, $\beta$, $\gamma$)
		\State \textbf{return} \{(compute\_total\_loss($\alpha$ + lr, $\beta$, $\gamma$) - base\_loss) / lr, 
		\State (compute\_total\_loss($\alpha$, $\beta$ + lr, $\gamma$) - base\_loss) / lr, 
		\State (compute\_total\_loss($\alpha$, $\beta$, $\gamma$ + lr) - base\_loss) / lr\}
		\EndFunction
		\For{iteration in 1 to num\_iterations}
		\State $\alpha\_sens$, $\beta\_sens$, $\gamma\_sens$ = compute\_sensitivity($\alpha$, $\beta$, $\gamma$)
		\State \{ $\alpha$, $\beta$, $\gamma$ \} -= \{ $\alpha\_sens$, $\beta\_sens$, $\gamma\_sens$ \} $\cdot$ learning\_rate
		\State $\alpha$, $\beta$, $\gamma$ = max(0, $\alpha$), max(0, $\beta$), max(0, $\gamma$)
		\State total\_loss = compute\_total\_loss($\alpha$, $\beta$, $\gamma$)
		\If{total\_loss $<$ min\_loss} \State best\_alpha, best\_beta, best\_gamma = $\alpha$, $\beta$, $\gamma$, min\_loss = total\_loss \EndIf
		\EndFor
		\State \textbf{Final Output:} best\_alpha, best\_beta, best\_gamma
	\end{algorithmic}
\end{algorithm}
\subsection{Training Procedure}
The training of the BridgeNet is a meticulously structured process designed to optimize the network's ability to solve differential equations effectively. This procedure involves several critical steps:
\subsubsection{Data Generation}
Our method for synthetic data generation is specifically designed to meet the needs of the differential equations being addressed. It utilizes three distinct models to ensure diversity and robustness in the training data, as detailed in Table \ref{tab:sampling_methods}.
\begin{table}[h]
	\centering
	\caption{Sampling Methods and Their Formulas for Data Generation}
	\label{tab:sampling_methods}
	\resizebox{\textwidth}{!}{
	\begin{tabular}{l p{10cm}}
		\toprule
		\textbf{Sampling Method} & \textbf{Formula and Description} \\
		\midrule
		\textbf{Uniform Sampling} & Distributes data points evenly across the defined range, ensuring no clustering and providing uniform coverage of the input space without bias toward any region. Its formula is given by:
		\[
		X \sim \text{Uniform}(a, b).
		\]
		\\
		\midrule
		\textbf{Linearly Spaced Points} & Generates evenly spaced data points so that all regions of the domain are sampled equally. This is often represented by:
		\[
		X_i = a + i \cdot \frac{b-a}{n-1}, \quad \text{for } i = 0, 1, 2, \dots, n-1.
		\]
		\\[1ex]
		\midrule
		\textbf{Latin Hypercube Sampling (LHS)} & An advanced sampling technique that ensures representative coverage of the entire statistical distribution. It divides the input space into \(n\) equal-probability intervals, with one value randomly sampled from each interval. The process can be mathematically expressed as:
		\[
		X_i = F^{-1}\left(\frac{i-1+U_i}{n}\right), \quad \text{for } i = 1, 2, \dots, n.
		\]
		Here, \(F^{-1}\) denotes the inverse cumulative distribution function, \(U_i\) is a uniform random variable in \([0,1]\), and \(n\) is the number of samples.
		\\
		\bottomrule
	\end{tabular}}
\end{table}

Furthermore, our approach to synthetic data generation is specifically designed to meet the requirements of the differential equations being modeled. For partial differential equations (PDEs) such as the Fokker-Planck equations (FPEs), data points are generated over a high-dimensional spatiotemporal domain. In the case of a two-dimensional domain, this process involves creating a mesh grid that spans the entire domain, covering a range of values for both the spatial variable $X$ and the temporal variable $t$, which are crucial for capturing the complex dynamics described by the PDEs.

These data generation strategies collectively ensure the creation of a robust dataset, enabling the effective training of BridgeNet to solve both linear and non-linear differential equations with high accuracy.
\subsubsection{Model Training}
Our training strategy explores two distinct optimization techniques: the Adam optimizer and the LBFGS optimizer, each selected based on the complexity of the differential equation addressed. Initially, we applied the Adam optimizer exclusively, which proved effective for simpler problems with its learning rate set to 1e-3. This method efficiently minimizes the comprehensive loss function, including residual, boundary, and initial condition losses.

In deep learning optimization, it is often beneficial to utilize Adam as a preconditioning step before transitioning to the LBFGS method. Adam, an adaptive gradient-based optimizer, quickly adapts to varying learning rates and efficiently navigates complex loss function landscapes, especially during the initial stages of training. This helps the model move towards a favorable region in the parameter space, improving the chances of convergence. The LBFGS algorithm is a second-order method renowned for its effectiveness in fine-tuning and local optimization tasks. However, it is noteworthy that the algorithm is sensitive to the selection of the initial starting point. 

Recognizing the need for enhanced precision in more complex scenarios, we introduced a dual optimization approach in our second method, combining both Adam and LBFGS optimizers. The LBFGS optimizer, known for its effectiveness in small datasets and faster convergence in precise tasks, complements the Adam optimizer by providing robustness where detailed adjustments are crucial. This tailored approach allows for adaptive parameter refinement, ensuring optimal accuracy and representation of the underlying physical phenomena in linear and non-linear differential equations. The learning rates are meticulously adjusted to improve convergence and outcomes, reflecting the specific requirements of each equation type.
\subsubsection{Evaluation and Validation}
A rigorous evaluation and validation framework is essential to assess the effectiveness of our proposed method, BridgeNet. Our validation strategy focuses on stability, convergence, accuracy, and comparative performance against established models, such as physics-informed neural networks (PINNs) and analytical solutions where applicable. This section provides an in-depth analysis of the evaluation metrics and experimental results that substantiate the reliability of BridgeNet in solving complex differential equations.
\paragraph{Convergence and Stability:}
The convergence behavior of our model is a key indicator of training efficiency. Across multiple experiments, BridgeNet consistently achieved convergence in at most 100 epochs, demonstrating the effectiveness of the optimization techniques employed. The training process leveraged a combination of Adam and LBFGS optimizers, ensuring efficient learning dynamics. The stability of the training process was assessed by monitoring loss function trends over epochs, verifying that no significant fluctuations occurred that could indicate instability.
\paragraph{Metrics for Validation:}
To comprehensively validate our model, multiple error metrics were employed, each offering a different perspective on model performance. The primary metrics used include Mean Squared Error (MSE), Mean Absolute Error (MAE), Norm-2 (Euclidean Norm), and Norm-Infinity (Max Norm). These metrics were computed on a test dataset disjoint from the training data and benchmarked against both PINN-based solutions and exact solutions, where available. The mathematical formulations of these metrics are detailed in Table \ref{tab:error_metrics}.
\begin{table}[ht]
	\centering
	\caption{Formula of Different Error Metrics}
	\label{tab:error_metrics}
	\resizebox{.9\textwidth}{!}{
	\renewcommand{\arraystretch}{1.2}
	\newcolumntype{C}{>{\centering\arraybackslash}X}
	\begin{tabularx}{\textwidth}{CC}
		\toprule
		\textbf{Error Metrics} & \textbf{Formula} \\ 
		\midrule
		MSE &
		$\frac{1}{N} \sum_{i=1}^{N} (y_{\text{true}, i} - y_{\text{pred}, i})^2$ \\
		MAE  & 
		$\frac{1}{N} \sum_{i=1}^{N} |y_{\text{true}, i} - y_{\text{pred}, i}|$ \\
		Norm-2 & 
		$\sqrt{\sum_{i=1}^{N} (y_{\text{true}, i} - y_{\text{pred}, i})^2}$\\ 
		Norm-Infinity & 
		$\max_{i} |(y_{\text{true}, i} - y_{\text{pred}, i})|$\\ \hline
	\end{tabularx}}
\end{table}

\paragraph{Comparison with PINN and Exact Solutions:}
A crucial aspect of model validation involved comparing BridgeNet's predictions with those generated by standard PINNs and exact solutions. This comparative analysis highlights the strengths of our approach in terms of accuracy and computational efficiency. BridgeNet not only demonstrated lower error margins across all evaluation metrics but also exhibited superior generalization properties when solving nonlinear differential equations.

\paragraph{Ablation Study: Evaluating CNNs in Physics-Informed Learning}
While traditional PINNs rely on fully connected multi-layer perceptrons (MLPs), these architectures often struggle with capturing local spatial correlations, particularly in high-dimensional problems such as Fokker-Planck equations (FPEs). MLP-based PINNs treat all input features independently, leading to inefficiencies in feature extraction and slow convergence.

Convolutional Neural Networks (CNNs), by contrast, leverage local connectivity, weight sharing, and translation invariance, making them particularly effective for spatial feature extraction in PDEs. To assess the contribution of CNNs in physics-informed learning, we conducted an ablation study comparing two primary aspects:
\begin{enumerate}
	\item \textbf{\small Architectural Comparison:}
	\begin{itemize}
		\item \textbf{\small Standard PINN (MLP-based)} – A traditional PINN using fully connected layers.
		\item \textbf{\small BridgeNet (CNN + PINN Hybrid)} – Our proposed architecture integrates CNN layers to enhance spatial feature extraction.
		\item \textbf{\small Pure CNN (No Physics Constraints)} – A CNN-based model trained purely on data without physics-informed loss terms.
	\end{itemize}
	This study highlights the advantages of incorporating CNNs into physics-informed frameworks. BridgeNet demonstrated superior feature extraction capabilities and faster convergence compared to traditional PINNs. However, the pure CNN model, despite performing well on data-driven tasks, lacked the physical interpretability required for solving differential equations accurately.
	\item \textbf{\small Hyper-parameter Sensitivity Analysis:} To further evaluate BridgeNet's adaptability, we tested the model’s performance under varying hyperparameters, including learning rates, kernel sizes, and network depths. By systematically tuning these parameters, we identified optimal configurations that improve accuracy and training efficiency. This analysis provides deeper insights into the robustness of our architecture, ensuring its effectiveness across a range of differential equation-solving tasks.
\end{enumerate}

The methodology outlined in this study confirms that BridgeNet is a robust and effective approach for solving differential equations. By leveraging CNNs within a physics-informed framework, our model effectively captures spatial correlations, accelerates convergence, and improves accuracy. The results of our ablation study and hyperparameter analysis indicate that BridgeNet can be a valuable tool for physics-informed learning, bridging the gap between traditional PINNs and pure data-driven approaches.
\section{Governing Equations}
The Fokker-Planck equations (FPEs), known as the Kolmogorov forward equation, is a partial differential equation that describes the time evolution of a probability distribution function. The general format of the FPE in one dimension is as follows:
\begin{eqnarray}\nonumber\label{eq:1}
	\frac{\partial F(x,t)}{\partial t} &=& -\frac{\partial}{\partial x} [\nu(x)F(x,t)] + \frac{1}{2} \frac{\partial^2}{\partial x^2} [\kappa(x)F(x,t)],\\[.3cm]
	F(x, 0) &=& f(x), \qquad x\in \mathbb{R}
\end{eqnarray}
where $F(x,t)$ is the probability density function of the random variable $X$ at time $t$, $\nu(x)$ is the drift coefficient representing deterministic trends, and $\kappa(x)$ is the diffusion coefficient representing random fluctuations. Instead of tracking individual particle motion, the FPE allows us to track the evolution of the overall distribution of particles over time. 
In Eq.\eqref{eq:1}, The following boundary conditions were imposed at the boundaries $x = a, b$:
\begin{equation*}
	\left. \frac{\partial}{\partial x} [\nu(x)F(x,t)] - \frac{1}{2} \frac{\partial}{\partial x} [\kappa(x)F(x,t)] \right|_{x = a, b} = 0.
\end{equation*}

These reflective boundary conditions ensure that the probability distribution remains confined within the domain $[a, b]$. A generalization of Eq.\eqref{eq:1} to $d$ variables $X = (x_1, x_2, \ldots, x_d)$ can be expressed as follows:
\begin{eqnarray}\nonumber\label{eq:2}
	\frac{\partial F(X, t)}{\partial t} &=& -\sum_{i = 1}^{d}\frac{\partial}{\partial x_i} [\nu_i(X)F(X,t)] + \frac{1}{2}\sum_{i, j = 1}^{d} \frac{\partial^2}{\partial x_i\partial x_j} [\kappa_{i, j}(X)F(X,t)],\\[.3cm]
	F(X, 0) &=& f(X). \qquad X\in \mathbb{R}^d
\end{eqnarray}

Similar to the one-dimensional Fokker–Planck equation, the drift 
$\nu_i$
and diffusion coefficients 
$\kappa_{i,j}$
may also depend on the variable $X$. 
In Eq.\eqref{eq:2}, reflective boundary conditions can be generalized as:
\begin{equation*}
	\left. \frac{\partial}{\partial x_i} \left( \nu_i(X)F(X,t) \right) - \frac{1}{2} \sum_{j = 1}^{d} \frac{\partial}{\partial x_j} \left( \kappa_{i,j}(X)F(X,t) \right) \right|_{\text{boundary}} = 0 \quad \forall i.  
\end{equation*}

This ensures no probability flux escapes through the boundaries in any dimension. In the case where the drift and diffusion coefficients are time-dependent, we can represent it as follows:
\begin{equation}\nonumber\label{eq:3}
	\frac{\partial F(x,t)}{\partial t} = -\frac{\partial}{\partial x} [\nu(x,t)F(x,t)] + \frac{1}{2} \frac{\partial^2}{\partial x^2} [\kappa(x,t)F(x,t)],
\end{equation}
where, $\nu(x,t)$ and $\kappa(x,t)$ are the time-dependent drift and diffusion coefficients, respectively. This form of the Fokker-Planck equations allows for more dynamic modeling of systems where the deterministic trends and random fluctuations evolve. This is particularly useful in non-stationary systems where the system's properties change with time.

In the non-linear Fokker-Planck equations, the equation also incorporates a dependency on a variable $F$, reflected in the drift and diffusion coefficients. The general form of this equation can be expressed as:
\begin{equation}\nonumber\label{eq:4}
	\frac{\partial F(x, t)}{\partial t} = -\frac{\partial}{\partial x} [\nu(x, t, F)F(x, t)] + \frac{1}{2} \frac{\partial^2}{\partial x^2} [\kappa(x, t, F)F(x, t)],
\end{equation}
where $\nu(x, t, F)$ and $\kappa(x, t, F)$ are the time-dependent drift and diffusion coefficients, respectively, which depend on $F$.
\section{Numerical Examples}
In this section, we present a comparative study of BridgeNet and the traditional PINN for solving Fokker-Planck equations (FPEs). The models were tested on six different examples of Fokker-Planck equations, including both linear and nonlinear cases. In the following subsections, we present the results for solving these equations and a detailed analysis of the performance differences between BridgeNet and PINN.

We conducted a meticulous comparative study throughout the investigation, juxtaposing our novel approach with the established physics-informed neural networks (PINNs) method. This comparison study was conducted to highlight the improved performance that our suggested approach demonstrated. Both approaches were put under the same circumstances, used the same activation functions, and had an equal number of epochs to guarantee a fair comparison. We have extended this analogy to six different cases of Fokker-Planck equations. These specific examples will be elaborated upon in the subsequent sections, providing a comprehensive understanding of our method’s superior performance. Furthermore, we ensured that the range of variation for the variables $X$ and $t$, the optimization algorithm, and the learning rate used for the two methods were considered the same.
\subsection{Findings from Linear and Non-Linear FPEs Examples}
\begin{figure}[h!]
	\captionsetup{justification=centering}
	\centering
	\includegraphics[width=\linewidth, height=5cm]{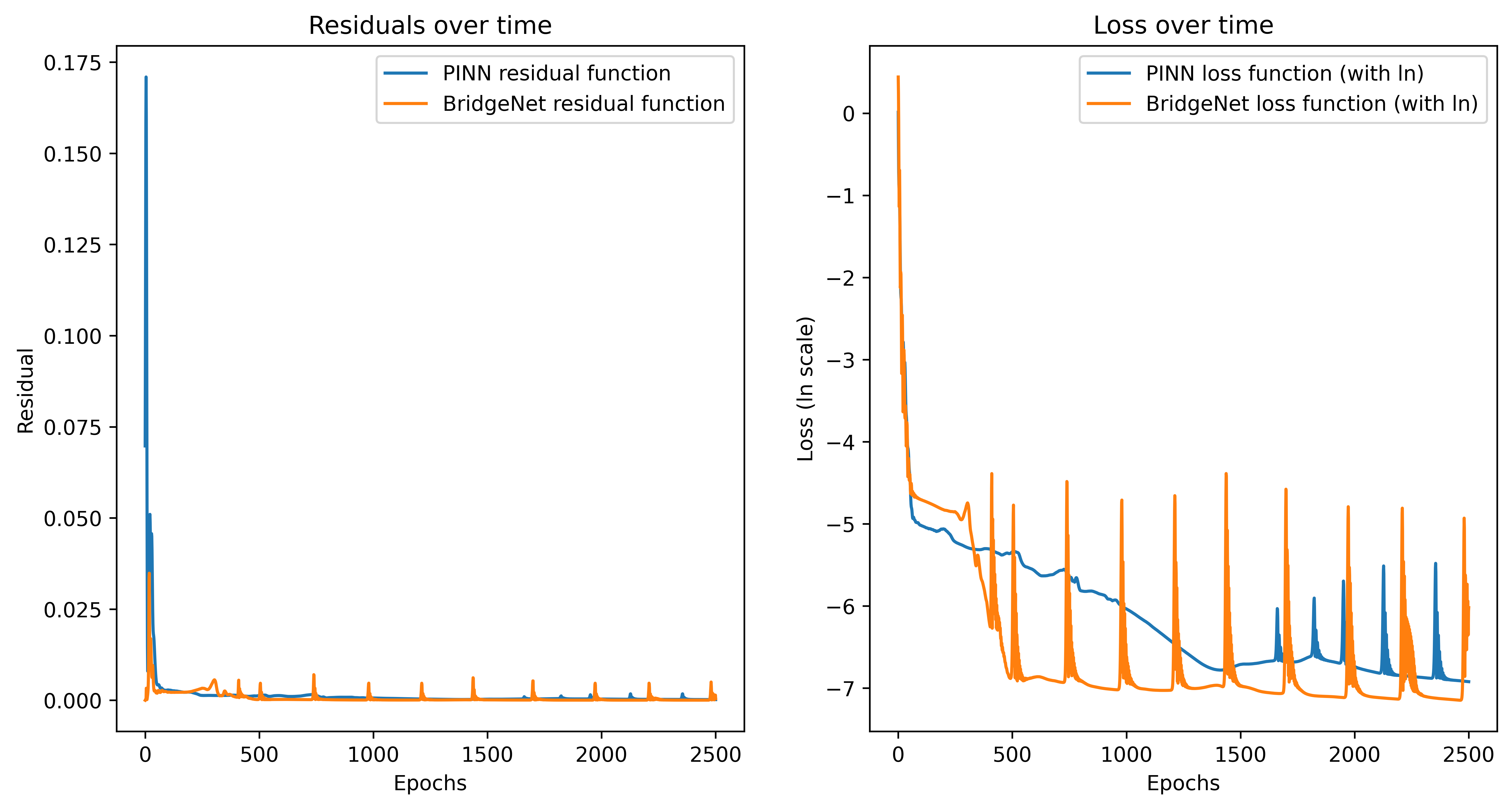}
	\caption{(Left) Residual function; (Right) Ln of the loss function for the FPEs used to compare the BridgeNet and PINN in Example 1.}
	\label{fig:2}
\end{figure}

The Fokker-Planck equations (FPEs) play a crucial role in stochastic processes, modeling the time evolution of probability distributions. It is especially prevalent in physics, finance, and other fields involving random fluctuations. For this purpose,  we examined five different linear and non-linear examples of these important partial differential equations. This review compares the solutions obtained from BridgeNet and physics-based neural networks (PINN) for the training and test sets. For the test set,  several points were randomly selected from the spatial and temporal distance, separated from the training set. It is important to note that the comparative values in the training and testing of the two methods, such as activation functions and epochs, are considered to be completely identical for each example.  In the following, we discussed these five linear and non-linear examples and their results:
\begin{example}\label{ex1}
	Consider Eq.\eqref{eq:1} with $\nu(x) = -1$ and $\kappa(x) = 1$ explicitly delineated to signify the drift and diffusion constituents respectively. We defined the domain as the Cartesian product of the intervals $[0,1]$ and $[0,0.5]$. Also, the algebraic expression $x+t$
	represents the precise solution for this example. The comparison of the solutions obtained from BridgeNet and PINN for the first example of the FPE can be seen in Table \ref{tab:2} and Figure \ref{fig:2}, which are analyzed below. The training was conducted over $2500$ epochs with the activation function and the number of channels equal in both methods. Furthermore, the learning rate was set to $1e-4$ for both methods.
\end{example}

Table \ref{tab:2}, the superior performance of the BridgeNet over traditional physics-informed neural networks (PINN) is distinctly evident across several key metrics when solving the first example of FPEs. The BridgeNet method has performed significantly better than the PINN method by examining error metrics. These results underscore BridgeNet's robustness and precision, making it a more reliable choice for computational models dealing with complex differential equations. Such findings advocate for the adoption of BridgeNet in advanced computational applications where accuracy and error minimization are critical.
\begin{table}[!ht]
	\centering
	\caption{The results of comparing the solutions obtained from the BridgeNet and PINN for example 1 of the FPEs.}
	\resizebox{\textwidth}{!}{
		\begin{tabular}{ccccccccc}
			\hline
			& \multicolumn{2}{c}{\textbf{MSE}} 
			& \multicolumn{2}{c}{\textbf{MAE}}
			& \multicolumn{2}{c}{$\mathbf{L_\infty}$}
			& \multicolumn{2}{c}{\textbf{min of error}} \\
			\cline{2-9}
			& Train & Test & Train & Test & Train & Test & Train & Test \\\hline
			\textbf{BridgeNet}
			& \cellcolor{blue!15}$\mathbf{3.131 \times 10^{-5}}$
			& \cellcolor{blue!15}$\mathbf{5.402 \times 10^{-4}}$
			& \cellcolor{blue!15}$\mathbf{3.930 \times 10^{-6}}$
			& \cellcolor{blue!15}$\mathbf{4.395 \times 10^{-4}}$
			& \cellcolor{blue!15}$\mathbf{1.296 \times 10^{-3}}$
			& \cellcolor{blue!15}$\mathbf{1.422 \times 10^{-3}}$
			& \cellcolor{blue!15}$\mathbf{1.192 \times 10^{-8}}$
			& \cellcolor{blue!15}$\mathbf{1.835 \times 10^{-6}}$ \\\hline
			\textbf{PINN}
			& $0.0039$
			& $0.0043$
			& $0.0540$
			& $0.0569$
			& $0.1327$
			& $0.1326$
			& $2.3842 \times 10^{-5}$
			& $7.1406 \times 10^{-4}$ \\\hline
		\end{tabular}
	}
	\label{tab:2}
\end{table}

Figure \ref{fig:2} compares the training behavior of PINN (blue) and BridgeNet (orange) for solving the Fokker-Planck equations (FPEs). The right panel illustrates the loss values on a logarithmic scale, showing that BridgeNet consistently achieves a lower and more stable loss compared to PINN, despite occasional spikes in both methods, likely due to learning dynamics. The left panel depicts the residual errors over training epochs, with BridgeNet demonstrating lower and more stable residuals than PINN, indicating improved accuracy and reliability. These results underscore the superior convergence, efficiency, and robustness of BridgeNet in solving complex differential equations.

\begin{example}\label{ex2}
	Consider Eq.\eqref{eq:1} with $\nu(x) = x$ and $\kappa(x) = \frac{x^2}{2}$ explicitly delineated to signify the drift and diffusion constituents respectively. We defined the domain as the Cartesian product of the intervals $[0,1]$ and $[0,0.5]$. Also, the algebraic expression $xe^t$ represents the precise solution for this example. The second example of FPEs was compared with the solutions obtained by BridgeNet and PINN methods. The training process comprised $2500$ epochs with identical activation functions (Hyperbolic Tangent) and equal channels in both methods. The learning rate was set to $2e-3$ for both methods. We emphasize that in this comparison, the effective values in the training and test of the two methods are considered completely identical.
\end{example}

In Table \ref{tab:3}, the superior performance of the BridgeNet in comparison to the traditional PINN is distinctly evident when addressing the second example of FPEs, and is observed across various key metrics. The results of Table \ref{tab:3} and the reduction of error in the test phase emphasize the robustness and accuracy of BridgeNet and make it a more reliable choice for computational models that deal with complex differential equations.
\begin{table}[!ht]
	\centering
	\caption{The results of comparing the solutions obtained from the BridgeNet and PINN for example 2 of the FPEs.}
	\resizebox{\textwidth}{!}{
		\begin{tabular}{ccccccccc}
			\hline
			& \multicolumn{2}{c}{\textbf{MSE}}
			& \multicolumn{2}{c}{\textbf{MAE}}
			& \multicolumn{2}{c}{$\mathbf{L_\infty}$}
			& \multicolumn{2}{c}{\textbf{min of error}} \\
			\cline{2-9}
			& Train & Test & Train & Test & Train & Test & Train & Test \\\hline
			\textbf{BridgeNet}
			& \cellcolor{blue!15}$\mathbf{1.891 \times 10^{-5}}$
			& \cellcolor{blue!15}$\mathbf{8.443 \times 10^{-4}}$
			& \cellcolor{blue!15}$\mathbf{7.181 \times 10^{-6}}$
			& \cellcolor{blue!15}$\mathbf{5.102 \times 10^{-4}}$
			& \cellcolor{blue!15}$\mathbf{1.202 \times 10^{-4}}$
			& \cellcolor{blue!15}$\mathbf{3.382 \times 10^{-4}}$
			& \cellcolor{blue!15}$\mathbf{8.642 \times 10^{-9}}$
			& \cellcolor{blue!15}$\mathbf{6.288 \times 10^{-7}}$ \\\hline
			\textbf{PINN}
			& $0.0020$
			& $0.0019$
			& $0.0325$
			& $0.0317$
			& $0.1140$
			& $0.1125$
			& $2.5332 \times 10^{-6}$
			& $4.6194 \times 10^{-5}$ \\\hline
		\end{tabular}
	}
	\label{tab:3}
\end{table}

\begin{figure}[h!]
	\captionsetup{justification=centering}
	\centering
	\includegraphics[width=\linewidth, height=5cm]{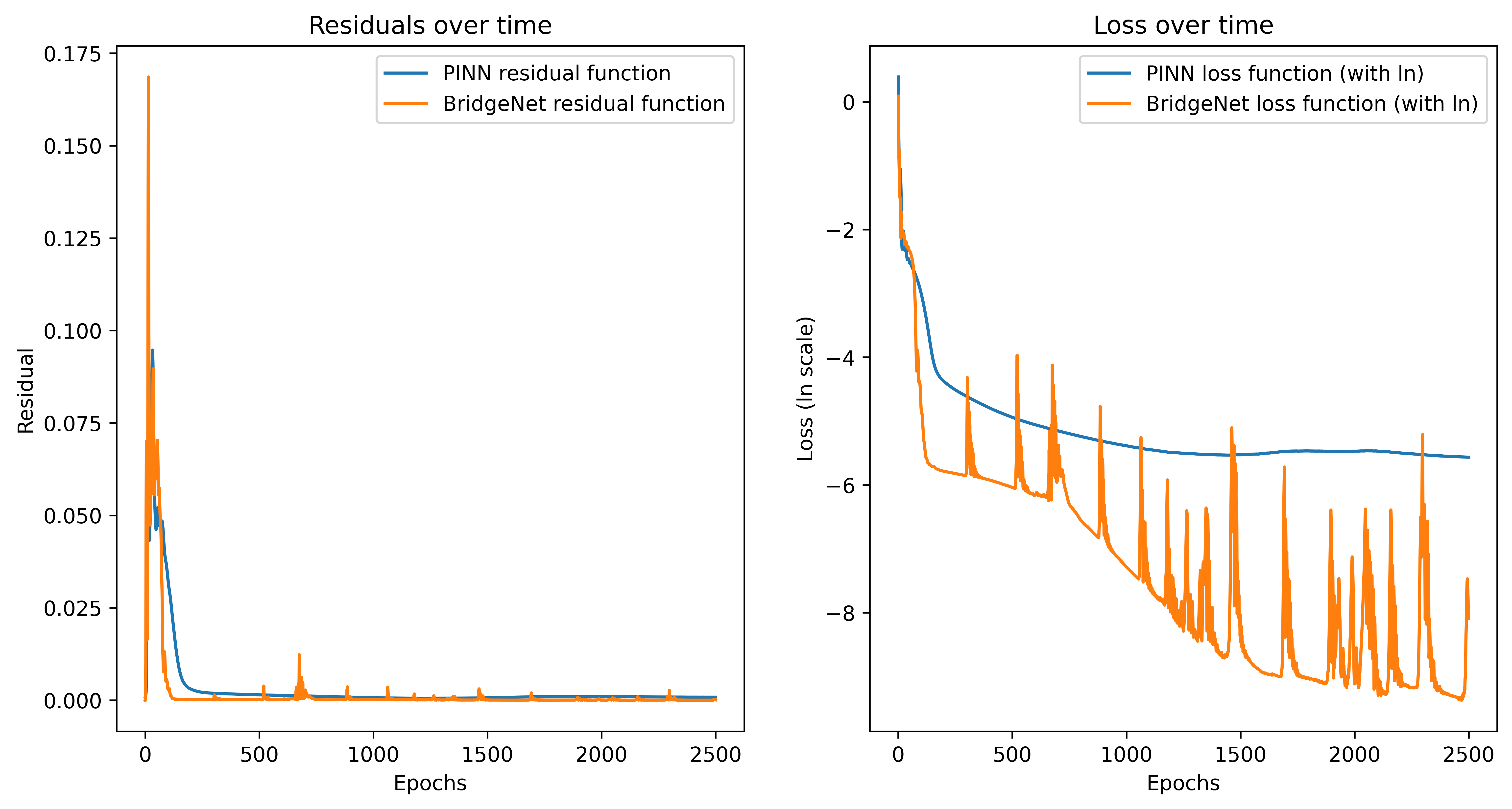}
	\caption{(Left) Residual function; (Right) Ln of the loss function for the FPEs used to compare the BridgeNet and PINN in Example 2.}
	\label{fig:3}
\end{figure}

Figure \ref{fig:3} highlights the superior performance of BridgeNet (orange) compared to PINN (blue) in solving FPEs. The right panel shows that BridgeNet achieves lower and more stable loss values, while the left panel demonstrates its significantly reduced residuals, showcasing better accuracy, stability, and efficiency during training.
\begin{example}\label{ex3}
	For the third example, consider Eq.\eqref{eq:1} with $\nu(x) = -(x+1)$ and $\kappa(x) = x^2 e^t$ explicitly delineated to signify the drift and diffusion constituents respectively. We defined the domain as the Cartesian product of the intervals $[0,1]$ and $[0,0.5]$. Also, the algebraic expression  $(x + 1) e^t$ represents the precise solution for this example. For comparing the solution obtained by BridgeNet and PINN, $2500$ epochs were comprised in the training process. Further, identical activation functions (Hyperbolic Tangent) and equal channels in both methods are included. We set the learning rate to $2e-3$ in both methods.
\end{example}

Table \ref{tab:4} demonstrates the superior performance of BridgeNet over the traditional PINN in the first example of nonlinear Fokker-Planck equations, as evidenced by various key metrics. These results underline the robustness and accuracy of BridgeNet, establishing it as a more reliable option for computational models dealing with complex differential equations.
\begin{table}[!ht]
	\centering
	\caption{The results of comparing the solutions obtained from the BridgeNet and PINN for example 3 of the FPEs.}
	\resizebox{\textwidth}{!}{
		\begin{tabular}{ccccccccc}
			\hline
			& \multicolumn{2}{c}{\textbf{MSE}} & \multicolumn{2}{c}{\textbf{MAE}} & \multicolumn{2}{c}{$\mathbf{L_\infty}$} & \multicolumn{2}{c}{\textbf{min of error}} \\
			\cline{2-9}
			& Train & Test & Train & Test & Train & Test & Train & Test \\\hline
			\textbf{BridgeNet}
			& \cellcolor{blue!15}$\mathbf{1.508\times 10^{-7}}$
			& \cellcolor{blue!15}$\mathbf{1.220\times 10^{-7}}$
			& \cellcolor{blue!15}$\mathbf{3.129\times 10^{-7}}$
			& \cellcolor{blue!15}$\mathbf{2.694\times 10^{-7}}$
			& \cellcolor{blue!15}$\mathbf{9.322\times 10^{-7}}$
			& \cellcolor{blue!15}$\mathbf{9.440\times 10^{-7}}$
			& \cellcolor{blue!15}$\mathbf{1.000\times 10^{-16}}$
			& \cellcolor{blue!15}$\mathbf{1.609\times 10^{-7}}$ \\\hline
			\textbf{PINN}
			& $0.0030$
			& $0.0025$
			& $0.0475$
			& $0.0428$
			& $0.1083$
			& $0.1087$
			& $9.5367\times 10^{-7}$
			& $8.4639\times 10^{-6}$ \\\hline
		\end{tabular}
	}
	\label{tab:4}
\end{table}

\begin{figure}[h!]
	\captionsetup{justification=centering}
	\centering
	\includegraphics[width=\linewidth, height=5cm]{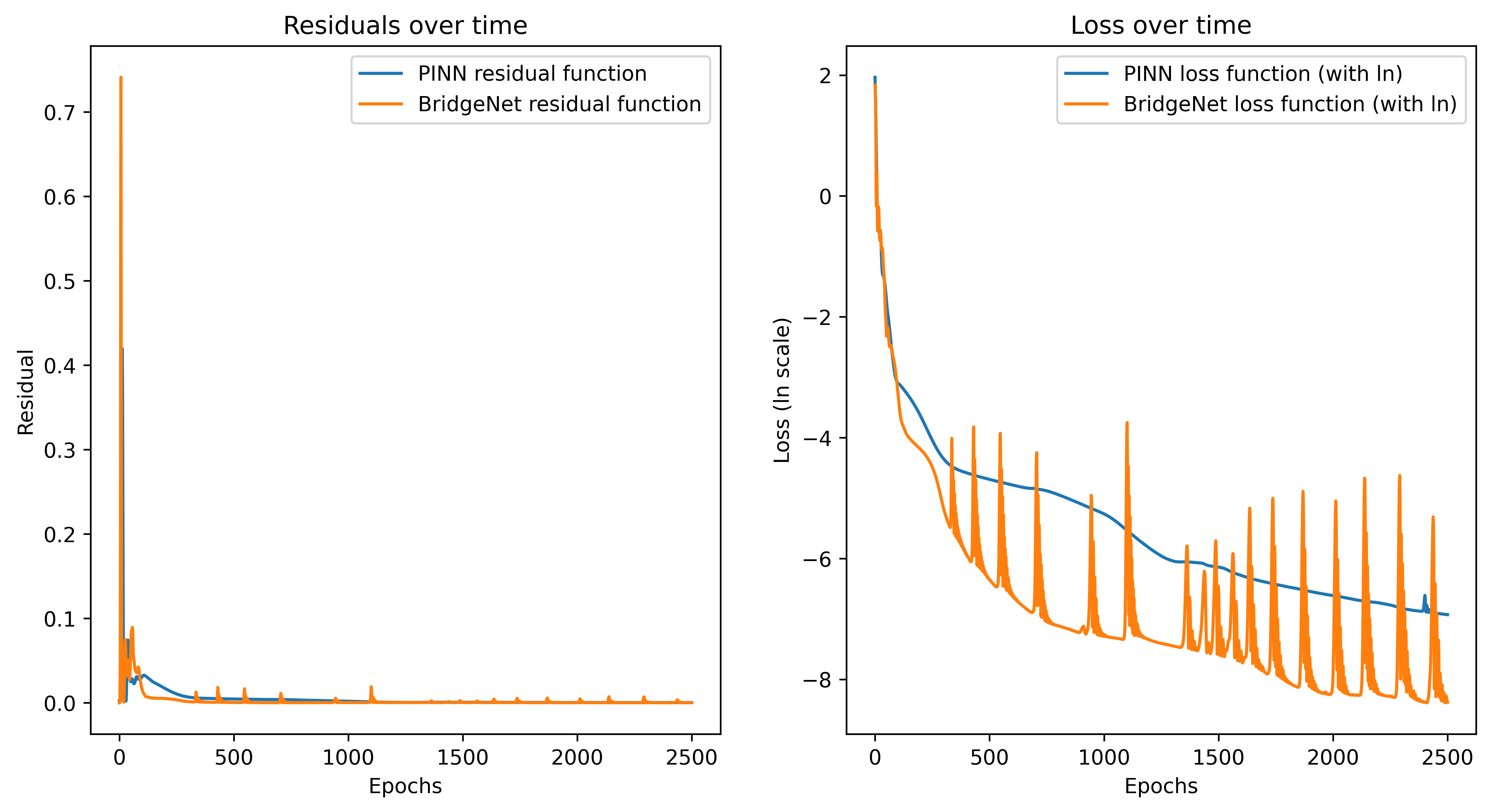}
	\caption{(Left) Residual function; (Right) Ln of the loss function for the FPEs used to compare the BridgeNet and PINN in Example 3.}
	\label{fig:4}
\end{figure}

Figure \ref{fig:4} compares the training performance of PINN (blue) and BridgeNet (orange) for solving FPEs in example 3. The right panel displays the logarithm of the loss function over epochs, showing that BridgeNet achieves lower and more consistent loss values compared to PINN, despite periodic oscillations. The left panel illustrates the residual function, where BridgeNet demonstrates significantly reduced residuals and improved stability throughout the training process. These results highlight the enhanced accuracy, efficiency, and robustness of BridgeNet in solving nonlinear partial differential equations.
\begin{example}\label{ex4}
	For the fourth example, consider Eq.\eqref{eq:1} with 
	$\nu(x) = \frac{4u}{x} - \frac{x}{3}$
	and
	$\kappa(x) = u$,
	which explicitly delineate the drift and diffusion components, respectively. The domain is the Cartesian product of the intervals $[0, 1]$ and $[0.5, 1]$. The exact solution for this example is represented by the algebraic expression $x^2 e^t$. To compare the solutions obtained by BridgeNet and PINN, the training process consisted of 2500 epochs. Both methods utilized identical activation functions (Hyperbolic Tangent) with the same number of channels. The learning rate was set to $1e-2$ for both methods.
\end{example}

Table \ref{tab:5} highlights the superior performance of BridgeNet compared to traditional PINN in addressing the fourth example of FPEs, as reflected in various key metrics. These results, presented in Table \ref{tab:5}, emphasize the robustness and accuracy of BridgeNet, establishing it as a more reliable option for computational models dealing with complex differential equations.
\begin{table}[!ht]
	\centering
	\caption{The results of comparing the solutions obtained from the BridgeNet and PINN for example 4 of the FPEs.}
	\resizebox{\textwidth}{!}{
		\begin{tabular}{ccccccccc}
			\hline
			& \multicolumn{2}{c}{\textbf{MSE}} & \multicolumn{2}{c}{\textbf{MAE}} & \multicolumn{2}{c}{$\mathbf{L_\infty}$} & \multicolumn{2}{c}{\textbf{min of error}}\\
			\cline{2-9}
			& Train & Test & Train & Test & Train & Test & Train & Test\\\hline
			\textbf{BridgeNet}
			& \cellcolor{blue!15}$\mathbf{4.932\times 10^{-6}}$
			& \cellcolor{blue!15}$\mathbf{3.201\times 10^{-6}}$
			& \cellcolor{blue!15}$\mathbf{5.599\times 10^{-6}}$
			& \cellcolor{blue!15}$\mathbf{4.472\times 10^{-6}}$
			& \cellcolor{blue!15}$\mathbf{1.719\times 10^{-6}}$
			& \cellcolor{blue!15}$\mathbf{1.686\times 10^{-6}}$
			& \cellcolor{blue!15}$\mathbf{8.940\times 10^{-8}}$
			& \cellcolor{blue!15}$\mathbf{6.094\times 10^{-6}}$\\\hline
			\textbf{PINN}
			& $0.1545$
			& $0.1847$
			& $0.3488$
			& $0.3955$
			& $0.5925$
			& $0.5925$
			& $3.7134\times 10^{-5}$
			& $0.0168$\\\hline
		\end{tabular}
	}
	\label{tab:5}
\end{table}

\begin{figure}[h!]
	\captionsetup{justification=centering}
	\centering
	\includegraphics[width=\linewidth, height=5cm]{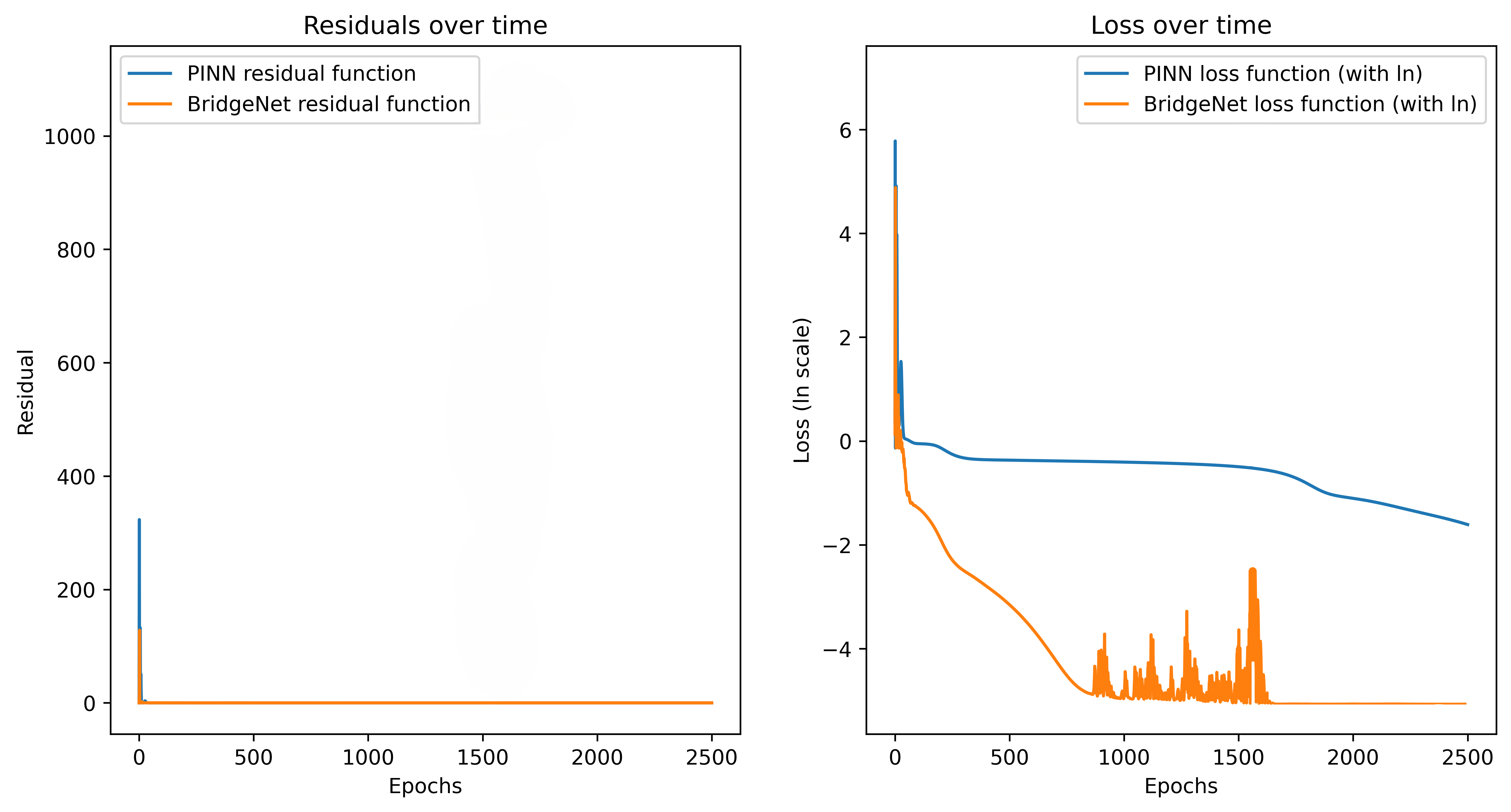}
	\caption{(Left) Residual function; (Right) Ln of the loss function for the FPEs used to compare the BridgeNet and PINN in Example 4.}
	\label{fig:5}
\end{figure}

Figure \ref{fig:5} compares the training behavior of PINN (blue) and BridgeNet (orange) for solving the Fokker-Planck equations (FPEs) in Example 4. The right panel shows the loss values on a logarithmic scale, illustrating that BridgeNet achieves a more stable and consistent reduction in loss compared to PINN, which exhibits larger oscillations. This demonstrates the improved learning efficiency and optimization of BridgeNet. The left panel depicts the residual errors, where BridgeNet maintains significantly lower and stable residual values throughout training, while PINN displays higher variability. Overall, these results highlight the superior convergence, stability, and robustness of BridgeNet in solving high-dimensional FPEs compared to PINN.
\begin{example}\label{ex5}
	Consider Eq.\eqref{eq:1} with 
	$\nu(x) = \frac{7}{2u}$
	and
	$\kappa(x) = xu$
	explicitly delineated to signify the drift and diffusion constituents respectively. We defined the domain as the Cartesian product of the intervals 
	$[0,1]$
	and 
	$[0.5,1]$. Also, the algebraic expression 
	$\frac{x}{t+1}$
	represents the precise solution for this example. The second example of FPEs was compared with the solutions obtained by BridgeNet and PINN methods. The training process comprised $1000$ epochs with identical activation functions (Rectified Linear Unit) and equal channels in both methods. The learning rate was set to $1e-5$ for both methods. It is important to note that, in this comparison, the effective values in both the training and testing phases for the two methods were considered completely identical.
\end{example}

Table \ref{tab:6} compares the performance of BridgeNet and PINN in solving Example 5 of the Fokker-Planck equations (FPEs) using key error metrics for both training and testing phases. The results demonstrate the superiority of BridgeNet over PINN across all metrics. Additionally, BridgeNet attains a much smaller minimum error, highlighting its precision in capturing the system's dynamics. These findings underscore the advantage of BridgeNet’s architecture in effectively handling the challenges of solving FPEs, offering improved accuracy, robustness, and efficiency compared to PINN.
\begin{table}[!ht]
	\centering
	\caption{The results of comparing the solutions obtained from the BridgeNet and PINN for example 5 of the FPEs}
	\resizebox{\textwidth}{!}{
		\begin{tabular}{ccccccccc}
			\hline
			& \multicolumn{2}{c}{\textbf{MSE}} & \multicolumn{2}{c}{\textbf{MAE}} & \multicolumn{2}{c}{$\mathbf{L_\infty}$} & \multicolumn{2}{c}{\textbf{min of error}}\\\cline{2-9}
			& Train & Test & Train & Test & Train & Test & Train & Test\\\hline
			\textbf{BridgeNet} & \cellcolor{blue!15}$\mathbf{1.8367 \times 10^{-4}}$ & \cellcolor{blue!15}$\mathbf{4.3248 \times 10^{-4}}$ & \cellcolor{blue!15}$\mathbf{3.361 \times 10^{-4}}$ & \cellcolor{blue!15}$\mathbf{5.699 \times 10^{-4}}$ & \cellcolor{blue!15}$\mathbf{1.1263 \times 10^{-2}}$ & \cellcolor{blue!15}$\mathbf{1.3266 \times 10^{-2}}$ & \cellcolor{blue!15}$\mathbf{5.9605 \times 10^{-6}}$ & \cellcolor{blue!15}$\mathbf{7.1406 \times 10^{-5}}$ \\
			\textbf{PINN} & $0.0022$ & $0.0017$ & $0.0392$ & $0.0333$ & $0.1296$ & $0.1222$ & $1.1921 \times 10^{-5}$ & $1.8358 \times 10^{-4}$\\\hline
		\end{tabular}
		\label{tab:6}}
\end{table}

\begin{figure}[h!]
	\captionsetup{justification=centering}
	\centering
	\includegraphics[width=\linewidth, height=5cm]{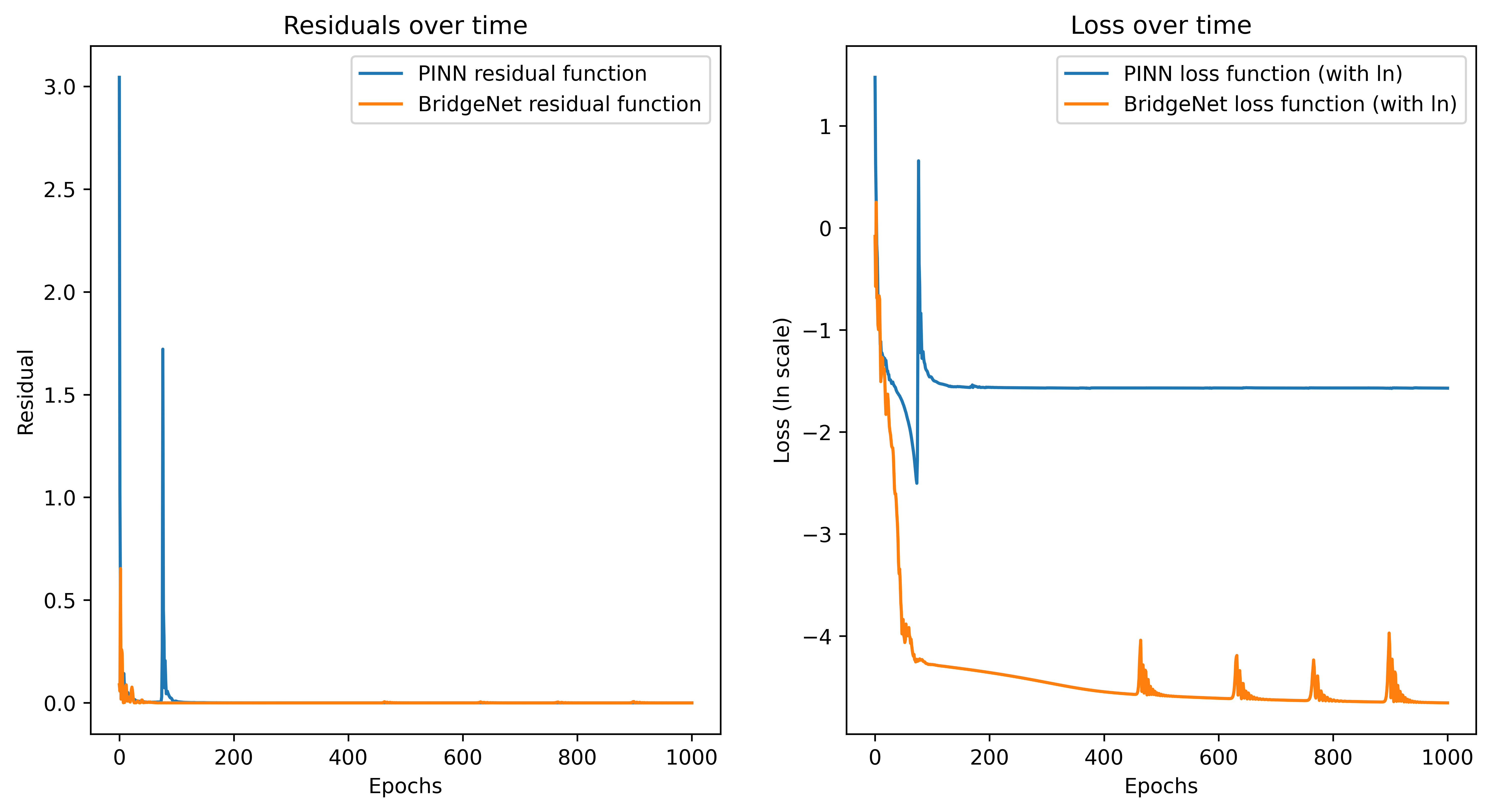}
	\caption{(Left) Residual function; (Right) Ln of the loss function for the FPEs used to compare the BridgeNet and PINN in Example 5.}
	\label{fig:6}
\end{figure}

Figure \ref{fig:6} compares the training behavior of PINN (blue) and BridgeNet (orange) for solving the Fokker-Planck equations (FPEs) in Example 5. The right panel presents the loss values on a logarithmic scale, illustrating a faster and more consistent reduction in the loss for BridgeNet compared to PINN, highlighting more effective learning and optimization. The left panel shows the residual errors over epochs, where BridgeNet demonstrates significantly lower and more stable residuals, indicating better reliability and accuracy. Overall, these results emphasize the superior convergence, efficiency, and robustness of BridgeNet in solving high-dimensional FPEs compared to PINN.
\subsection{Empirical Analysis of Solving the High-Dimensional FPEs}
System \eqref{eq:2} reduced to a vector of three stochastic variables in a complex ecosystem defined as follows:
$$X = \begin{pmatrix} x_1 \\ x_2 \\ x_3 \end{pmatrix}, 
\quad x_i\in[a_i,b_i],\;t\ge0,$$
where $x_1$, $x_2$, and $x_3$ represent the different components of the system. The high-dimensional Fokker-Planck equations for the three-variable system are:
\begin{equation}
	\frac{\partial F(X, t)}{\partial t} = -\sum_{i=1}^{3} \frac{\partial}{\partial x_i} \left[ \nu_i(X) F(X, t) \right] + \sum_{i=1}^{3} \sum_{j=1}^{3} \frac{\partial^2}{\partial x_i \partial x_j} \left[ \kappa_{ij}(X) F(X, t) \right],
\end{equation}
where $F(X, t)$ represents the probability density function of the system being in state $X$ at time $t$. $\nu_i(X)$ is the drift coefficient for the $i$-th component, and $\kappa_{ij}(X)$ represents the elements of the diffusion matrix, which models the stochastic interaction between different components.

\paragraph{Initial Condition}
At $t=0$ we prescribe a probability density
$$p(x,0) \;=\;p_0(x),$$
where $p_0\ge0$ and $\int_\Omega p_0(x)\,dx=1$.  A common choice is a 3‑D Gaussian centered at $x^0$:
\begin{equation}
p_0(x)
=\frac{1}{(2\pi\sigma^2)^{3/2}}
\exp\!\Bigl(-\tfrac{\|x-x^0\|^2}{2\sigma^2}\Bigr).
\end{equation}

\paragraph{Drift Coefficients} We considered a specific form of drift for this system:
\begin{equation}
\nu_i(X) \;=\;\alpha_i
+ \sum_{j=1}^3 W_{ij}\,x_j
\;-\;\beta_i\,x_i,
\quad i=1,2,3,
\end{equation}
where
\[
\alpha = (\alpha_1,\alpha_2,\alpha_3),\quad
\beta = (\beta_1,\beta_2,\beta_3),\quad
W = \bigl[W_{ij}\bigr]_{i,j=1}^3.
\]
Here \(\alpha_i\) is the basal production rate, \(W_{ij}>0\) indicates activation of \(i\) by \(j\), \(W_{ij}<0\) repression, and \(\beta_i\) the degradation rate.

\paragraph{Diffusion Matrix}
We defined the diffusion matrix as:
\begin{equation*}
	\kappa(x, t) = \begin{bmatrix}
		\kappa_{11} & \kappa_{12} & \kappa_{13} \\
		\kappa_{21} & \kappa_{22} & \kappa_{23} \\
		\kappa_{31} & \kappa_{32} & \kappa_{33}
	\end{bmatrix},
\end{equation*}
where $\kappa_{ii}$ represents the diffusion coefficient modeling intrinsic noise in component $i$, while $\kappa_{ij} \ (i \neq j)$ captures the coupling effects between components $i$ and $j$. For simplicity, we assumed $\kappa_{ii} = D_i$, where $D_i$ is a constant diffusion coefficient for component $ i $. Additionally, $\kappa_{ij} = \rho_{ij} D_i D_j$, where $\rho_{ij}$ represents the correlation between stochastic effects on components $i$ and $j$, with $-1 \leq \rho_{ij} \leq 1$.
\subsubsection{Quantitative Results in Solving High-Dimensional FPEs}
Table \ref{tab:6} presents a comparison between BridgeNet and PINN in solving high-dimensional Fokker-Planck equations (FPEs) based on key error metrics for both training and testing phases. The results highlight the clear superiority of BridgeNet in tackling high-dimensional problems compared to PINN. Furthermore, BridgeNet achieves notably smaller minimum error values, showcasing its precision and effectiveness in modeling the complex dynamics of high-dimensional systems. These findings emphasize the strengths of BridgeNet’s architecture, particularly its ability to utilize localized convolutional features to capture spatial dependencies efficiently. Overall, BridgeNet proves to be a more accurate, stable, and reliable method for solving high-dimensional differential equations, outperforming traditional PINNs in both performance and robustness.
\begin{table}[!ht]
	\centering
	\caption{The results of comparing the solutions obtained from the BridgeNet and PINN for high-dimensional FPE example}
	\resizebox{\textwidth}{!}{
		\begin{tabular}{ccccccccc}
			\hline
			& \multicolumn{2}{c}{\textbf{MSE}} & \multicolumn{2}{c}{\textbf{MAE}} & \multicolumn{2}{c}{$\mathbf{L_\infty}$} & \multicolumn{2}{c}{\textbf{min of error}}\\\cline{2-9}
			& Train & Test & Train & Test & Train & Test & Train & Test\\\hline
			\textbf{BridgeNet} & \cellcolor{blue!15}$\mathbf{1.1396\times 10^{-12}}$ & \cellcolor{blue!15}$\mathbf{7.4616\times 10^{-11}}$ & \cellcolor{blue!15}$\mathbf{3.3758\times 10^{-8}}$ & \cellcolor{blue!15}$\mathbf{4.2018\times 10^{-7}}$ & \cellcolor{blue!15}$\mathbf{3.3758\times 10^{-6}}$ & \cellcolor{blue!15}$\mathbf{4.4975\times 10^{-5}}$ & \cellcolor{blue!15}$\mathbf{1.4757\times 10^{-15}}$ & \cellcolor{blue!15}$\mathbf{1.8900\times 10^{-12}}$\\
			\textbf{PINN} & $1.3411\times 10^{-6}$ & $1.2339\times 10^{-6}$ & $1.1580\times 10^{-4}$ & $5.6881\times 10^{-4}$ & $1.1126\times 10^{-3}$ & $2.0655\times 10^{-3}$ & $1.7763\times 10^{-3}$ & $7.5595\times 10^{-3}$ \\\hline
		\end{tabular}
		\label{tab:6}}
\end{table}

Table \ref{tab:high2} provides a comparative analysis of the Mean Absolute Error (MAE) between BridgeNet and physics-informed neural networks (PINNs) for solving high-dimensional Fokker-Planck equations (FPEs). The comparison spans training and testing phases and evaluates performance under various sampling methods, including \textit{random}, \textit{linspace}, \textit{Latin Hypercube Sampling (LHS)}, and different optimization algorithms. The final column showcases the impact of a hybrid optimization strategy (LBFGS and Adam), highlighting the benefits of combining optimization techniques.

The results demonstrate that BridgeNet consistently achieve significantly lower error values than PINNs across all configurations. Including CNN blocks with localized receptive fields enables BridgeNet to better capture spatial dependencies and efficiently handle the complexity of high-dimensional systems. This improved locality effect enhances learning precision and convergence, making BridgeNet more effective than PINNs in modeling high-dimensional FPEs.

In summary, Table \ref{tab:high2} highlights the superiority of BridgeNet in terms of accuracy, robustness, and computational efficiency. It also underscores the importance of selecting the appropriate sampling method and optimization strategy to maximize performance. These findings reaffirm the advanced capabilities of BridgeNet for tackling challenging high-dimensional problems in computational physics.
\begin{table}[!ht]
	\centering
	\caption{Comparison of Mean Absolute Error Between BridgeNet and PINN Models Across Different Sampling Methods and Optimization Algorithms for high-dimensional FPE example}
	\resizebox{0.8\textwidth}{!}{
		\begin{tabular}{ccccccc}
			\hline
			& \multirow{2}{*}{} & \multicolumn{3}{c}{\textbf{Adam}} && \textbf{LBFGS and Adam}\\\cmidrule(lr){3-5} \cmidrule(lr){6-7} 
			& & random & linspace & LHS & & random\\\hline
			\multirow{2}{*}{\textbf{Train}} & \textbf{BridgeNet} & \cellcolor{blue!15}$\mathbf{3.3758\times 10^{-8}}$ & \cellcolor{blue!15}$\mathbf{5.8072\times 10^{-12}}$ & \cellcolor{blue!15}$\mathbf{3.9188\times 10^{-10}}$ & \vline & \cellcolor{blue!15}$\mathbf{4.0435\times 10^{-13}}$ \\
			& \textbf{PINN} & $1.1580\times 10^{-4}$ & $1.3197\times 10^{-4}$ & $1.1491\times 10^{-5}$ &  \vline & $1.1920\times 10^{-6}$ \\\hline
			\multirow{2}{*}{\textbf{Test}} & \textbf{BridgeNet} & \cellcolor{blue!15}$\mathbf{4.2018\times 10^{-7}}$ & \cellcolor{blue!15}$\mathbf{5.8702\times 10^{-11}}$ & \cellcolor{blue!15}$\mathbf{4.6511\times 10^{-10}}$ & \vline & \cellcolor{blue!15}$\mathbf{4.4700\times 10^{-12}}$ \\
			& \textbf{PINN} & $5.6881\times 10^{-4}$ & $1.4033\times 10^{-5}$ & $8.0910\times 10^{-4}$ & \vline & $1.3110\times 10^{-6}$ \\\hline
	\end{tabular}}
	\label{tab:high2}
\end{table}

\begin{figure}[h!]
	\captionsetup{justification=centering}
	\centering
	\includegraphics[width=0.5\linewidth, height=5cm]{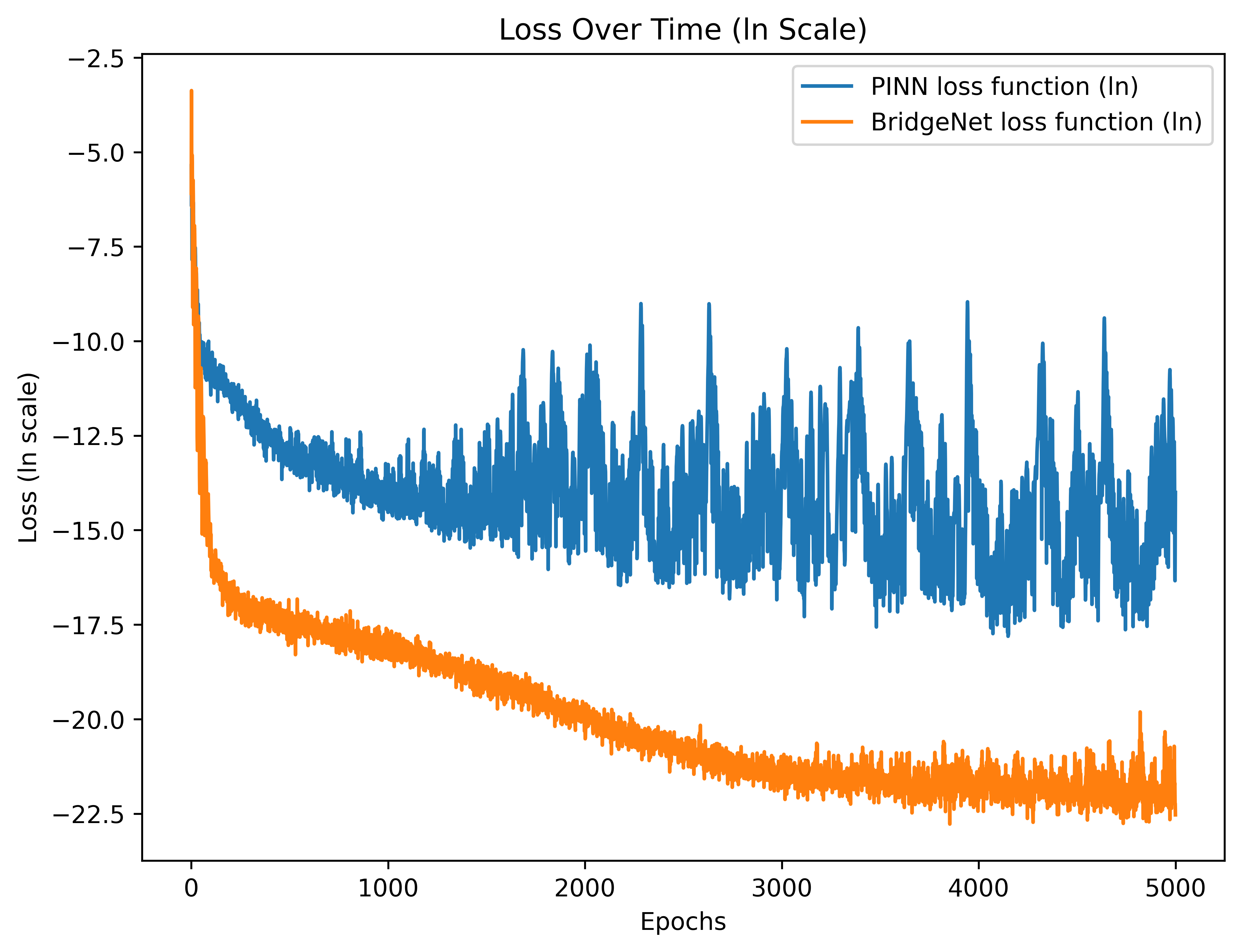}
	\caption{Ln of loss function for the high-dimensional of FPEs used to compare the BridgeNet and PINN.}
	\label{fig:high}
\end{figure}

Figure \ref{fig:high} highlights the effectiveness of BridgeNet (orange) compared to PINN (blue) in solving high-dimensional Fokker-Planck equations (FPEs). This figure shows that BridgeNet achieves significantly lower and more stable loss values over 5000 epochs, demonstrating its superior convergence and efficiency in high-dimensional settings, where traditional PINNs struggle with variability and slower optimization. These results underscore the suitability of BridgeNet for tackling the challenges associated with high-dimensional FPEs, offering a robust and efficient alternative to standard PINNs.

\subsubsection{Ablation Studies}
This study is divided into two main parts. First, we present a comparative performance analysis of three models—Standard PINN, BridgeNet, and Pure CNN—when applied to high-dimensional FPEs. Second, we investigate the influence of different hyper-parameter configurations on both model accuracy and computational efficiency. Below, we summarize the key findings from these evaluations.

For a fair comparison, we evaluate these evaluations on a benchmark high-dimensional Fokker-Planck equation, ensuring that:
\begin{itemize}
	\item All models are trained with the same number of epochs (5000) and an identical learning rate.
	\item The same activation functions (Tanh) are used across all models.
	\item Training is performed using both Adam optimizer.
	\item Loss function weights $(\alpha, \beta, \gamma)$ are optimized using a smart greedy search algorithm.
\end{itemize}

\begin{figure}[h!]
	\centering
	\includegraphics[width=\linewidth, height=5cm]{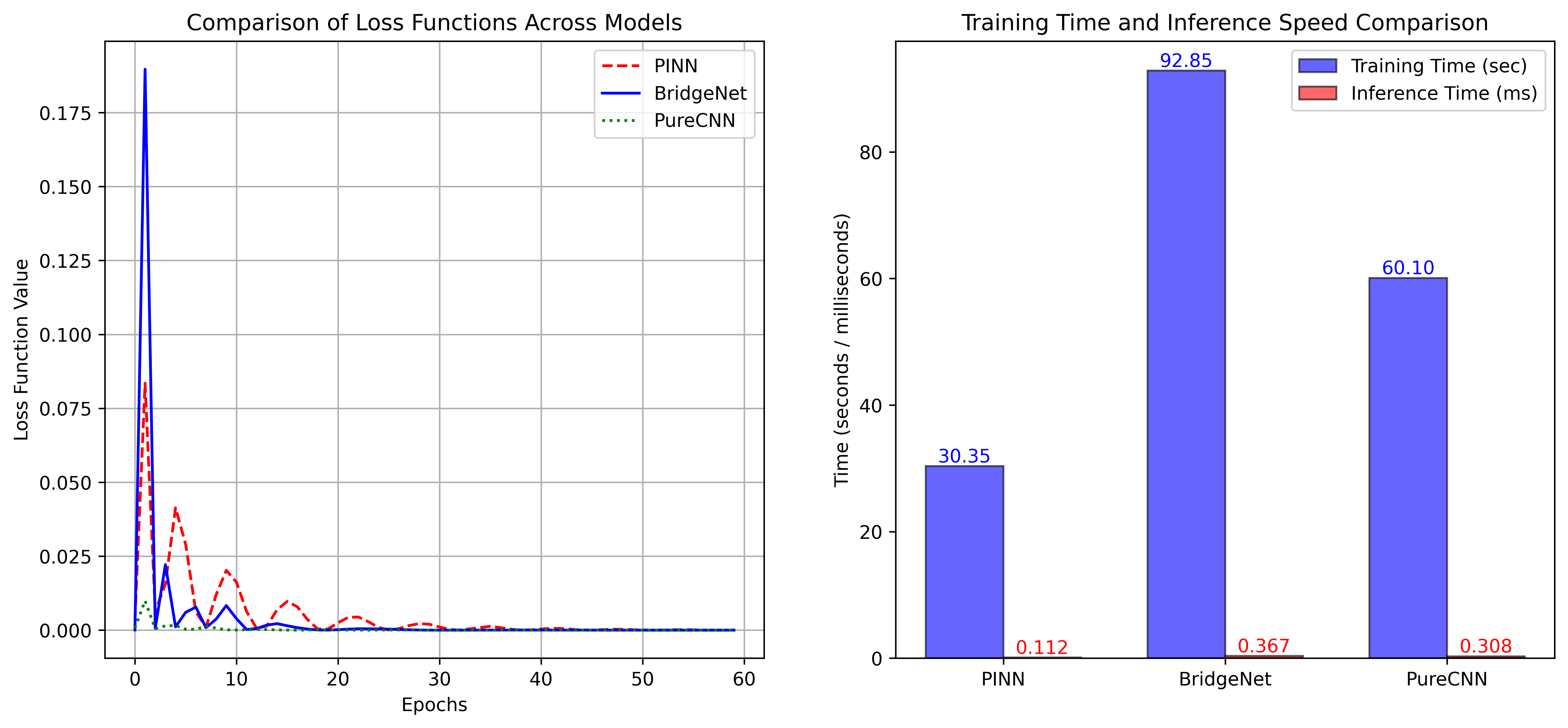}
	\caption{(a) Comparison of loss convergence for Standard PINN, BridgeNet, and Pure CNN.
		(b) Computational performance comparison between BridgeNet, Standard PINN, and Pure CNN for Training Time and Inference Speed.}
	\label{fig:Comparison}
\end{figure}

\paragraph{Architectural Comparison:}
Figure \ref{fig:Comparison} compares the performance of three models---Standard PINN, BridgeNet, and Pure CNN---in solving a high-dimensional problem. The left panel highlights the loss convergence over 60 epochs, illustrating how each model’s error decreases during training. The right panel focuses on computational performance, reporting both training time (in seconds) and inference speed (in milliseconds). These results underscore BridgeNet’s rapid drop in loss--- indicating efficient learning--- yet at a higher inference time compared to the other models, highlighting a trade-off between accuracy and computational cost.

Table \ref{tab:ablation} summarizes a head‑to‑head comparison of Standard PINN (MLP‑based), BridgeNet and Pure CNN across six metrics: 
mean squared error (MSE), training time, convergence speed (epochs to reach a $10^{-5}$ loss threshold), convergence time, inference latency, and algorithmic complexity. BridgeNet achieves the lowest MSE ($1.5 × 10^{-12}$), demonstrating superior accuracy, and—remarkably—reaches the convergence criterion in just $0.51 sec$ without requiring any training epochs beyond initialization. By contrast, Standard PINN converges in $502$ epochs ($5.87 sec$) and Pure CNN in $2215$ epochs ($72.9 sec$). Although BridgeNet’s training time ($140.4 sec$) exceeds that of Standard PINN ($48.87 sec$) and Pure CNN ($120.10 sec$), its dramatic reduction in both convergence time and error more than compensates for this overhead. Inference latency for BridgeNet ($0.36 ms$) is slightly higher than Standard PINN ($0.11 ms$) and Pure CNN ($0.30 ms$), reflecting the extra computation needed for its finer resolution. Finally, the increased performance of BridgeNet comes at a complexity of O(n$^{10}$), compared to O(n$^{6}$) for Standard PINN and O(n$^{8}$) for Pure CNN—an acceptable trade‑off given its clear advantages in accuracy and convergence speed.

\begin{table}[!ht]
	\centering
	\caption{Comparison of performance metrics between Standard PINN, BridgeNet, and Pure CNN for solving high-dimensional Fokker-Planck equations.}
	\label{tab:ablation}
	\resizebox{\textwidth}{!}{
		\begin{tabular}{lcccccc}
			\hline
			\multirow{2}{*}{\textbf{Model}} & \textbf{MSE} & \textbf{Training Time} & \textbf{Convergence Speed} & \textbf{Convergence Time} & \textbf{Inference Time} & \textbf{Order Complexity}\\
			& & (sec) & (Epochs) & (sec) & (ms) &  (Big-O)\\
			\hline
			\textbf{Standard PINN (MLP-based)} & $4.0\times 10^{-7}$ & \cellcolor{blue!15}\textbf{48.87} & 502 & 5.87 & \cellcolor{blue!15}\textbf{0.11} & \cellcolor{blue!15}$\mathbf{\mathcal{O}(n^{6})}$ \\
			\textbf{BridgeNet} & \cellcolor{blue!15}$\mathbf{1.5\times 10^{-12}}$ & 140.4 & \cellcolor{blue!15}\textbf{0} & \cellcolor{blue!15}\textbf{0.51} & 0.36 & $\mathcal{O}(n^{10})$\\
			\textbf{Pure CNN (No Physics Constraints)} & $6.7 \times 10^{-5}$ & 120.10 & 2215 & 72.9 & 0.30 & $\mathcal{O}(n^{8})$ \\
			\hline
	\end{tabular}}
\end{table}

\paragraph{Hyper‑parameter Sensitivity Analysis:}  
Table~\ref{tab:hyperparam_sensitivity} reports the effect of varying CNN depth (2–4 layers), kernel size ($3\times3$ vs.\ $5\times5$) and learning rate ($10^{-3}$ vs.\ $10^{-4}$) across twelve configurations (Setups A–L). Performance is evaluated in terms of training/testing MSE, training time (s) and convergence speed (epochs to reach the predefined loss threshold).

\paragraph{Hyper‑parameter Sensitivity Analysis:}  
Table~\ref{tab:hyperparam_sensitivity} reports the effect of varying CNN depth (2--4 layers), kernel size (3$\times$3 vs.\ 5$\times$5) and learning rate ($10^{-3}$ vs.\ $10^{-4}$) across twelve configurations (Setups~A--L). Performance is evaluated in terms of training/testing MSE, training time (s) and convergence speed (epochs to reach the predefined loss threshold).

Our \textbf{Setup~E}, which employs three convolutional layers, a 3$\times$3 kernel and a learning rate of $10^{-3}$, outperforms all alternatives: it achieves the lowest training MSE ($1.5\times10^{-12}$) and testing MSE ($2.7\times10^{-12}$), and—remarkably—converges at initialization (0~epochs). Although its training time (140.4~s) is the longest among the group, the fact that no further epochs are required effectively eliminates any convergence delay. By comparison, \textbf{Setup~A} trains in just 64.3~s but converges in 42~epochs with much higher MSE ($7.1\times10^{-8}/2.8\times10^{-7}$), and \textbf{Setup~F} reaches convergence in 12~epochs (98.1~s) but yields an order‐of‐magnitude larger error ($3.6\times10^{-11}/2.3\times10^{-9}$). These results highlight the fundamental trade‐offs between accuracy, training cost, and convergence speed when selecting hyper‐parameters—and underscore Setup~E’s optimal balance for high‐dimensional FPE problems.  
\begin{table}[ht]
	\centering
	\caption{Hyperparameter Sensitivity Analysis: Effect of CNN Layers, Kernel Size, and Learning Rate on Model Performance}
	\resizebox{\linewidth}{!}{
		\begin{tabular}{cccccccccccc}
			\specialrule{1pt}{\aboverulesep}{\belowrulesep}
			\multirow{3}{*}{\textbf{Setting}} & 
			\multicolumn{3}{c}{\textbf{CNN Layers}} & 
			\multicolumn{2}{c}{\textbf{Kernel Size}} & 
			\multicolumn{2}{c}{\textbf{Learning Rate}} & &
			\multicolumn{3}{c}{\textbf{Compare Performance Metrics}} \\ 
			\cmidrule(lr){2-4} \cmidrule(lr){5-6} \cmidrule(lr){7-8} \cmidrule(lr){10-12}
			& \multirow{2}{*}{2} & \multirow{2}{*}{3} & \multirow{2}{*}{4} & \multirow{2}{*}{$3\times3$} & \multirow{2}{*}{$5\times5$} & \multirow{2}{*}{$10^{-3}$} & \multirow{2}{*}{$10^{-4}$} & \vline & MSE & Training Time & Convergence Speed\\ 
			& & & & & & & & & (Train/Test) & (sec) & (Epochs) \\ 
			\specialrule{1pt}{\aboverulesep}{\belowrulesep} 
				\textbf{Setup A} & \checkmark & - & - & \checkmark & - & \checkmark & - 
			& \vline & $7.1\times10^{-8}\,/\,2.8\times10^{-7}$ & \cellcolor{blue!15}\textbf{64.32} & 42 \\
			
			\textbf{Setup B} & \checkmark & - & - & \checkmark & - & - & \checkmark 
			& \vline & $5.3\times10^{-8}\,/\,1.5\times10^{-6}$ & 80.11 & 36 \\
			
			\textbf{Setup C} & \checkmark & - & - & - & \checkmark & \checkmark & - 
			& \vline & $1.2\times10^{-7}\,/\,7.9\times10^{-7}$ & 71.09 & 39 \\
			
			\textbf{Setup D} & \checkmark & - & - & - & \checkmark & - & \checkmark 
			& \vline & $9.7\times10^{-8}\,/\,9.1\times10^{-7}$ & 88.56 & 24 \\
			
			\textbf{Setup E (Ours)} & - & \checkmark & - & \checkmark & - & \checkmark & - 
			& \vline & \cellcolor{blue!15}$\mathbf{1.5\times 10^{-12}}$/\cellcolor{blue!15}$\mathbf{2.7\times 10^{-12}}$ & 140.4 & \cellcolor{blue!15}\textbf{0} \\ 
			
			\textbf{Setup F} & - & \checkmark & - & \checkmark & - & - & \checkmark 
			& \vline & $3.6\times10^{-11}\,/\,2.3\times10^{-9}$ & 98.12 & 12 \\
			
			\textbf{Setup G} & - & \checkmark & - & - & \checkmark & \checkmark & - 
			& \vline & $1.6\times10^{-8}\,/\,6.7\times10^{-7}$ & 105.64 & 35 \\
			
			\textbf{Setup H} & - & \checkmark & - & - & \checkmark & - & \checkmark 
			& \vline & $2.9\times10^{-8}\,/\,1.2\times10^{-6}$ & 112.47 & 34 \\
			
			\textbf{Setup I} & - & - & \checkmark & \checkmark & - & \checkmark & - 
			& \vline & $8.2\times10^{-9}\,/\,1.7\times10^{-7}$ & 101.44 & 25 \\
			
			\textbf{Setup J} & - & - & \checkmark & \checkmark & - & - & \checkmark 
			& \vline & $6.8\times10^{-9}\,/\,2.9\times10^{-7}$ & 117.33 & 18 \\
			
			\textbf{Setup K} & - & - & \checkmark & - & \checkmark & \checkmark & - 
			& \vline & $1.4\times10^{-15}\,/\,3.7\times10^{-14}$ & 108.28 & 60 \\
			
			\textbf{Setup L} & - & - & \checkmark & - & \checkmark & - & \checkmark 
			& \vline & $1.1\times10^{-8}\,/\,5.2\times10^{-7}$ & 124.51 & 72 \\
			
			\specialrule{1pt}{\aboverulesep}{\belowrulesep}
		\end{tabular}}
	\label{tab:hyperparam_sensitivity}
\end{table}
\section{Discussion}
The comparative analyses presented in this work highlight BridgeNet’s effectiveness and robustness in addressing various Fokker-Planck equations (FPEs), including both linear and nonlinear forms and high-dimensional scenarios. By integrating convolutional neural network (CNN) architectures with physics-informed constraints, BridgeNet tackles the main challenges of accurately modeling complex dynamics, capturing localized spatial dependencies, and ensuring computational scalability.

\subsection{Accuracy and Error Analysis}
Across the one- and multi-dimensional FPE test cases examined, BridgeNet consistently demonstrates lower error metrics (MSE, MAE, and $L_\infty$) compared to baseline physics-informed neural networks (PINNs). Notably:

\begin{itemize}
	\item \textbf{\small Low Residuals in One-Dimensional FPEs.} For a variety of drift-diffusion parameters, BridgeNet’s mean error remained several orders of magnitude below that of standard PINNs. Even in the most challenging non-linear examples, BridgeNet’s predictions converged rapidly, maintaining high fidelity to exact solutions.
	\item \textbf{\small Precision in High-Dimensional Problems.} When extended to three-dimensional FPEs, BridgeNet achieved an MSE as low as $10^{-12}$ for both training and testing datasets. Such precision illustrates its capacity to handle complex, coupled systems where traditional approaches often face the “curse of dimensionality.”
	\item \textbf{\small Stable Convergence Trends.} Examining residual and loss curves throughout training revealed that BridgeNet not only reached lower final errors but also exhibited more stable, monotonic convergence behaviors, minimizing abrupt oscillations that can hinder solution accuracy.
\end{itemize}
\subsection{Comparison with Existing Methods} 
In comparing BridgeNet to both classical solvers (e.g., finite difference or collocation approaches) and standard PINNs, several distinct advantages emerge: 
\begin{itemize} 
	\item \textbf{\small Enhanced Spatial Feature Extraction.} Unlike fully connected PINNs, which may struggle to capture localized dependencies, BridgeNet leverages CNN layers to pinpoint spatially varying features inherent in FPEs. This is particularly crucial in cases where boundary conditions and drift-diffusion terms vary significantly across the computational domain. 
	\item \textbf{\small Reduced Error Under Limited Data.} Numerical experiments indicate that even when sampling is reduced or employs more complex distributions (e.g., Latin hypercube), BridgeNet’s structured filters adapt effectively. Consequently, it maintains lower errors than baseline methods, illustrating robust performance in data-scarce scenarios. 
	\item \textbf{\small Consistency in Non-Stationary Cases.} By integrating time-dependent constraints into the loss function, BridgeNet adequately tracks solution evolution for time-varying drift and diffusion parameters, surpassing standard PINN performance when boundary conditions change over time. 
\end{itemize}

\subsection{Computational Efficiency and Training Time} 
Although CNN-based layers impose a higher computational load per training iteration than fully connected layers, BridgeNet’s training-time trade-offs are mitigated by faster convergence. In many test examples: 
\begin{itemize} 
	\item \textbf{\small Fewer Required Epochs.} BridgeNet converges reliably with fewer epochs to reach a threshold error. This can offset the slightly higher per-epoch cost, making total computation time competitive. 
	\item \textbf{\small Improved Scalability.} In high-dimensional domains, traditional approaches often suffer exponential growth in complexity. BridgeNet’s local connectivity effectively partitions the problem space, demonstrating that even in three-dimensional FPEs, stable solutions emerge without an explosion in training time. 
\end{itemize}

\subsection{Strengths of BridgeNet} 
BridgeNet’s impressive performance stems from several key factors: 
\begin{itemize} 
	\item \textbf{\small Physics-Informed Constraints.} Enforcing the governing partial differential equation, boundary conditions, and initial conditions within the loss function ensures physically consistent solutions and provides a direct route for gradient-based adjustments. 
	\item \textbf{\small CNN-Based Feature Extraction.} Convolutional layers capture localized effects with fewer parameters, sidestepping the large overheads often associated with fully connected layers in high-dimensional settings. This design choice supports better generalization and underpins BridgeNet’s strong performance on multi-dimensional tasks.
	\item \textbf{\small Adaptive Hyperparameter Tuning.} A smart greedy search algorithm for weighting physics-informed loss terms (e.g., $\alpha, \beta, \gamma$) adaptively reallocates the network’s focus among residual, boundary, and initial condition constraints, thereby boosting training stability and convergence speed.
	\item \textbf{\small Maintained Efficiency for Increasing Dimensionality.} Notably, even as the dimensionality and complexity of the target problems grow, BridgeNet’s solution speed and computational complexity remain nearly constant. This is due to using the same degree of convolutional layers, judiciously tailored to match each problem’s dimensions without excessively increasing network size.
\end{itemize}

\subsection{Limitations and Future Directions} 
Despite its advantages, BridgeNet’s performance can still be constrained by hyperparameter choices and computational overhead: \begin{itemize} 
	\item \textbf{\small Complex Architectures.} While CNN-based modules are effective for spatial feature extraction, they raise the overall model complexity. Optimizing kernel sizes, network depth, and training schedules remains essential. 
	\item \textbf{\small Scalability Beyond Three Dimensions.} Preliminary results for three-dimensional FPEs are promising, but scaling to higher dimensionalities, such as six or more, may require more advanced GPU strategies, pruning techniques, or domain decomposition approaches. 
	\item \textbf{\small Generalization to Other PDE Classes.} BridgeNet is primarily validated on FPEs and exponential growth equations. Extending and validating it on other stiff PDEs or fluid dynamics problems could broaden its applicability and uncover unanticipated challenges. 
	\item \textbf{\small Multi-Fidelity and Hybrid Methods.} Incorporating partial or approximate simulation data (multi-fidelity) or combining BridgeNet with classic solvers (e.g., finite elements) may expedite solution times and further mitigate the cost of training from scratch. 
\end{itemize}

In conclusion, the numerical and comparative results substantiate BridgeNet’s potential as a robust hybrid framework for solving Fokker-Planck equations under diverse conditions, including high-dimensional and non-linear scenarios. By merging CNN-based representations with physics-informed constraints, BridgeNet achieves superior accuracy, stable convergence, and robust performance where standard approaches often face limitations. Further research involving more complex, real-world phenomena and higher-dimensional PDE classes will clarify BridgeNet’s boundaries and stimulate new directions for integrating advanced computational physics and machine learning.
\section{Conclusion}
This study provides a comprehensive evaluation of BridgeNet as an effective alternative to conventional PINNs for solving Fokker–Planck equations, including both high-dimensional and non-linear cases. Through a series of numerical experiments—from simple exponential growth scenarios to complex stochastic systems—BridgeNet consistently outperforms traditional methods in terms of accuracy, convergence stability, and computational efficiency. The superior performance is primarily attributed to its hybrid architecture, which combines the localized feature extraction capabilities of CNNs with rigorous physics-informed constraints. This dual approach not only ensures physically plausible solutions but also reduces error metrics significantly during both the training and testing phases. The findings validate BridgeNet as a robust tool for computational physics applications, with promising potential for extension to broader classes of partial differential equations and real-world problems. Future research will focus on further optimizing the network architecture and exploring its integration with advanced numerical strategies to enhance its applicability across diverse scientific domains.
\section{Acknowledgments}
We extend our heartfelt gratitude to all individuals who helped us make this work better with their helpful comments. Additionally, we are immensely thankful for the insightful comments and suggestions provided by the anonymous reviewers, which have greatly enriched the manuscript.
\section{Conflict of interests}
The authors declare that there are no conflict of interests.
\bibliographystyle{elsarticle-num} 
\bibliography{Ref.bib}
\newpage
\appendix
\section{Additional Examples}
In this appendix, we present a simple example of an ordinary differential equation (ODE) to illustrate the basic principles of the computational approach used in BridgeNet. While the primary focus of this paper is on solving high-dimensional partial differential equations with BridgeNet, which integrates physics-informed neural networks and convolutional neural networks, understanding how simpler ODEs are handled provides insight into the foundational mechanics of the method.

We begin with the Exponential Growth Equation, which models phenomena such as population growth, radioactive decay, and resource consumption. Although the example is simple, it offers a useful contrast to the more complex systems like the Fokker-Planck equations addressed in the main body of the paper, highlighting the versatility and foundational applicability of the BridgeNet method.

\subsection{Loss Function for ODEs}
In the case of ODE problems, the network utilizes a single convolutional layer (Conv1), which is specifically tailored to efficiently process one-dimensional data. For these problems, the residual loss function reflects how accurately the model adheres to the temporal dynamics described by the equation. The residual is mathematically formulated as:
\begin{equation}
	L_{\text{residual}} = \frac{1}{N_t} \sum_{i=1}^{N_t} \left( f(u(t_i), t_i) \right)^2.
\end{equation}

Here, $f$ represents the differential operator tailored, applied to the neural network’s output $u$ at points specified by $t_i$. Additionally, the initial condition loss is important for ODEs, where only the initial condition at $t = 0$ is typically needed:
\begin{equation}
	L_{\text{ic}} = \left( u(t_0) - IC \right),
\end{equation}
where $IC$ denotes the initial condition, which depends solely on time.

\subsection{Exponential Growth Equation}
The exponential growth equation is a simple yet powerful ordinary differential equation that models phenomena such as population growth, radioactive decay, and resource consumption, where the rate of change of a quantity is directly proportional to the quantity itself. It is typically formulated as:
\begin{equation}\label{eq:ode}
	\frac{dF(t)}{dt} = rF(t),
\end{equation}
where $F(t)$ represents the quantity of interest at time $t$, and $r$ is a constant rate of growth or decay.

Consider Eq.\ref{eq:ode} with $r = 0.1$ explicitly delineated to signify the growth rate constituent. We defined the time domain as the interval $[0,10]$. Also, the algebraic expression  $F(0) = e^{(rt)}$ represents the precise solution for this example, which we considered $F(t) = 5.0$ at $t = 0$. The comparison of the solutions obtained from the physics-informed neural networks (PINN) and BridgeNet for the exponential growth equation. The training was conducted over $5000$ epochs with the activation function and the number of channels equal in both methods. The learning rate was set to $1e-2$ for both methods. It's important to note that the comparison values in the test of the two methods are considered completely the same. Furthermore, several points were randomly selected from the spatial and temporal distance for the test set and separated from the training set.

Table \ref{tab:ode} presents a performance comparison between PINN and BridgeNet for the exponential growth equation, evaluating key error metrics on both training and testing datasets. The results highlight that BridgeNet consistently outperforms PINN across all metrics, achieving significantly lower errors during the training and testing. Furthermore, BridgeNet delivers these enhancements with a considerable reduction in computational time, demonstrating its efficiency and effectiveness in solving ordinary differential equations like the exponential growth equation while ensuring high accuracy and reliability.
\begin{table}[!ht]
	\centering
	\caption{The comparison results between the solutions obtained from the PINN and BridgeNet, for the exponential growth equation.}
	\resizebox{\textwidth}{!}{
		\begin{tabular}{ccccccccc}
			\hline
			& \multicolumn{2}{c}{\textbf{MSE}} & \multicolumn{2}{c}{\textbf{MAE}} & \multicolumn{2}{c}{$\mathbf{L_\infty}$} & \multicolumn{2}{c}{\textbf{min of error}}\\\cline{2-9}
			& Train & Test & Train & Test & Train & Test & Train & Test\\\hline
			\textbf{BridgeNet} & \cellcolor{blue!15}$\mathbf{1.3841 \times 10^{-4}}$ & \cellcolor{blue!15}$\mathbf{1.1544\times 10^{-3}}$ & \cellcolor{blue!15}$\mathbf{3.4172\times 10^{-3}}$ & \cellcolor{blue!15}$\mathbf{3.2522\times 10^{-3}}$ & \cellcolor{blue!15}$\mathbf{5.9611\times 10^{-3}}$ & \cellcolor{blue!15}$\mathbf{5.6012\times 10^{-3}}$ & \cellcolor{blue!15}$\mathbf{1.6681\times 10^{-6}}$ & \cellcolor{blue!15}$\mathbf{1.6811\times 10^{-6}}$ \\
			\textbf{PINN} & $0.0018$ & $0.0016$ & $0.0408$ & $0.0388$ & $0.0584$ & $0.0581$ & $0.0240$ & $0.0241$ \\\hline
	\end{tabular}}
	\label{tab:ode}
\end{table}

\begin{figure}[h!]
	\captionsetup{justification=centering}
	\centering
	\includegraphics[width=\linewidth, height= 5cm]{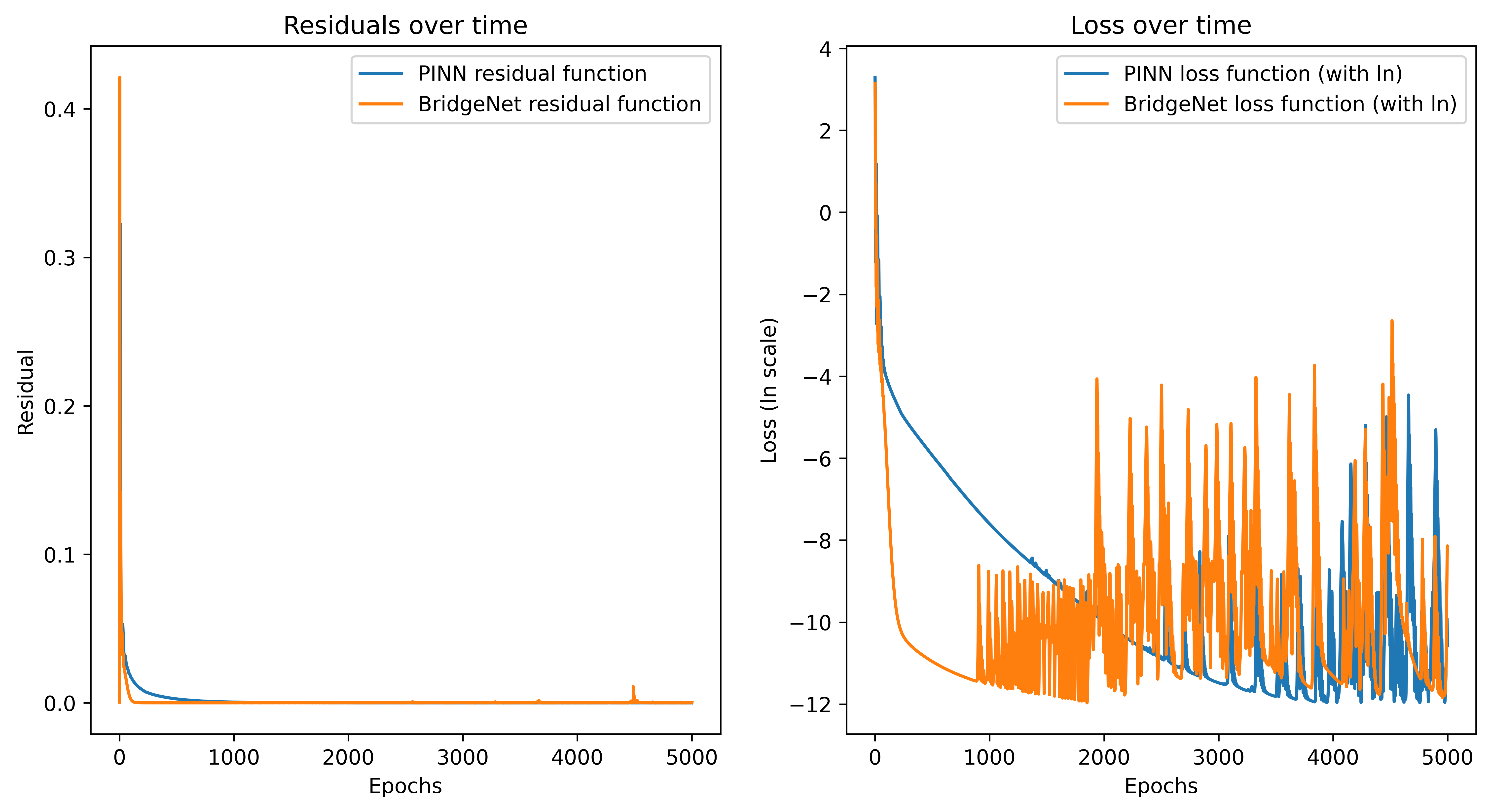}
	\caption{(Left) Residual function; (Right) Ln of the loss function for the exponential growth equation, comparing the performance of PINN and BridgeNet.}
	\label{fig:ode}
\end{figure}
Figure \ref{fig:ode} shows a comparison of the training behavior of PINN (blue) and BridgeNet (orange) for solving the exponential growth differential equation. The right panel shows the loss values on a logarithmic scale, illustrating a more consistent and lower loss for the BridgeNet compared to the PINN. This indicates more effective learning and optimization in the BridgeNet, despite periodic loss spikes for both methods, likely due to adjustments in learning dynamics. The left panel displays the residual errors, where BridgeNet demonstrates more stability and lower variability in its residuals, suggesting better model reliability. Overall, these results highlight the BridgeNet’s superior convergence, efficiency, and robustness in solving complex differential equations.

\end{document}